\documentclass[10pt]{IEEEtran}

\usepackage{flushend}
\usepackage[T1]{fontenc}
\usepackage{hyperref}
\hypersetup{colorlinks = true,	citecolor = {magenta},	urlcolor = {black}}
\usepackage{varwidth}
\usepackage{enumitem}
\usepackage{float}
\usepackage{graphicx}
\usepackage{rotating}
\usepackage{caption}
\usepackage[font=scriptsize,format=plain,labelfont=bf,textfont=bf]{subcaption}
\usepackage[font=scriptsize,format=plain]{caption}
\usepackage{wrapfig}
\usepackage{subcaption}
\usepackage{amsfonts}
\usepackage{multirow}
\usepackage{listings}
\usepackage{mdwlist}
\usepackage{comment}
\usepackage{scrextend}
\usepackage[linesnumbered,ruled,vlined]{algorithm2e}
\usepackage{cite}
\SetKw{KwBy}{by}
\usepackage{relsize}
\usepackage{tabularx}

\usepackage{bold-extra}
\usepackage{xcolor}
\usepackage[section]{placeins}
\usepackage{tablefootnote}
\usepackage[english]{babel}
\usepackage{hyperref}

\SetCommentSty{mycommfont}
\newcommand{\smartparagraph}[1]{\vspace{.05in}\noindent{\bf #1}}
\usepackage{mdwlist}
\usepackage{makecell}

\newcommand{\ap}[1]{\textit{\textcolor{purple}{[Apoorv]: #1}}}

\renewcommand{\footnotesize}{\scriptsize}
\usepackage{cleveref}
\crefformat{section}{\S#2#1#3} 
\crefformat{subsection}{\S#2#1#3}
\crefformat{subsubsection}{\S#2#1#3}

\newcommand{\prot}{{\textit{Verify}}\xspace}
\newcommand{\system}{$\textsc{Pazz}$\xspace}
\newcommand{\fuzz}{{Fuzzer}\xspace}
\newcommand{\consistencytester}{{Consistency Tester}\xspace}
\newcommand{\consistencychecker}{{Consistency Tester}\xspace}
\newcommand{\Consistencychecker}{{Consistency Tester}\xspace}
\newcommand{\ct}{{Consistency Tester}\xspace}
\newcommand{\cp}{{CPC}\xspace}
\newcommand{\dpl}{{DPC}\xspace}
\newcommand{\vt}{{\textit{VerifyTagType}}\xspace}
\newcommand{\vp}{{\textit{Verify\_Port}}\xspace}
\newcommand{\vr}{{\textit{Verify\_Rule}}\xspace}
\newcommand{\up}{{\textit{u\textsubscript{p}}}\xspace}
\newcommand{\ur}{{\textit{u\textsubscript{r}}}\xspace}
\newcommand{\vpd}{{\textit{Verify\_Port}}\xspace}
\newcommand{\vrd}{{\textit{Verify\_Rule}}\xspace}
\newcommand{\vtd}{{\textit{VerifyTagType}}\xspace}
\newcommand{\svr}{{\textit{set\_Verify\_Rule}}\xspace}
\newcommand{\svp}{{\textit{set\_Verify\_Port}}\xspace}

\begin{document}
\title{Consistent SDNs through Network State Fuzzing}

		\author{
		{\rm Apoorv Shukla$^1$ \quad S. Jawad Saidi$^2$ \quad Stefan Schmid$^3$ \quad Marco Canini$^4$ \quad Thomas Zinner$^1$ \quad Anja Feldmann$^{2,5}$ }\\
		\footnotesize TU Berlin$^1$ \quad MPI-Informatics$^2$  \quad University of Vienna$^3$ \quad KAUST$^4$ \quad UDS$^5$\\
	} 
\IEEEtitleabstractindextext{%
\begin{abstract}
The conventional wisdom is that a software-defined network (SDN) operates under the premise that the logically centralized control plane has an accurate representation of the actual data plane state.
Unfortunately, bugs, misconfigurations, faults or attacks can introduce inconsistencies that undermine correct operation.
Previous work in this area, however, lacks a holistic methodology to tackle this problem and thus, addresses only certain parts of the problem.
Yet, the consistency of the overall system is only as good as its least consistent part.

Motivated by an analogy of network consistency checking with program testing,
we propose to add active probe-based network state fuzzing to
our consistency check repertoire.  Hereby, our system, $\system$,
combines
production traffic with active probes to periodically test if the actual forwarding path and
decision elements (on the data plane) correspond to the expected ones (on the
control plane). Our insight is that active traffic covers the inconsistency cases beyond the ones identified by passive traffic. \system prototype was built and evaluated on topologies of varying scale and complexity. Our results show that \system requires minimal network resources to detect persistent data plane faults through fuzzing and localize them quickly while outperforming baseline approaches.
\end{abstract}
	
	}
\maketitle

\IEEEdisplaynontitleabstractindextext

%
%

\section{Introduction}
\label{sec:intro}

The correctness of a software-defined network (SDN) crucially depends on the consistency between the
management, the control and the data plane. There are, however, many causes that may trigger
inconsistencies at run time, including,
switch hardware failures~\cite{Survey, Zeng2014}, bit flips~\cite{bitflip, perevsini2015monocle,bitflip1,bitflip2,bitflip3,bitflip4}, misconfigurations~\cite{understandingBGP, dow, coned}, priority bugs~\cite{kuzniar2015you, cacheflow}, control and switch software bugs~\cite{jrex, soft, fib-ack}.
When an inconsistency occurs, the actual data plane state does not correspond to what the control plane expects it to be. Even worse, a malicious user may actively try to trigger inconsistencies as part of an attack vector.

\begin{figure}[tbh]
	\centering
	\includegraphics[width=\columnwidth]{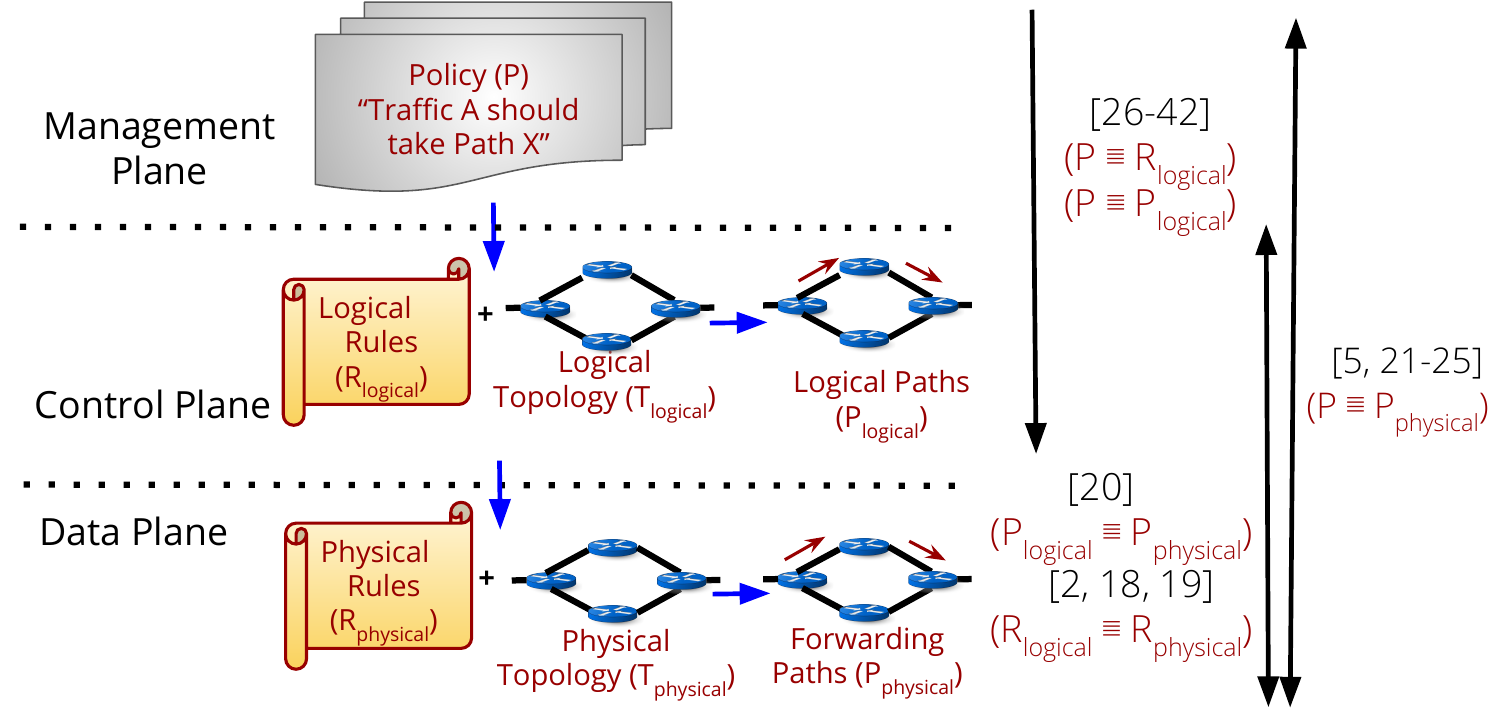}
	\captionof{figure}{Overview of consistency checks described in the literature.\label{fig:Rel}}
\end{figure}

Figure~\ref{fig:Rel} shows a visualization inspired by the one by Heller et al.~\cite{Heller2013} highlighting where consistency checks operate. The figure illustrates
the three network planes -- management, control, and data plane -- with their
components.  The management plane establishes the network-wide policy $P$, which
corresponds to the network operator's intent. To realize this policy, the control plane governs a set
of logical rules ($R$\textsubscript{logical}) over a logical topology ($T$\textsubscript{logical}), which yield a set of logical paths ($P$\textsubscript{logical}).
The data plane consists of the actual topology ($T$\textsubscript{physical}), the rules ($R$\textsubscript{physical}), and the resulting forwarding paths ($P$\textsubscript{physical}).

Consistency checking is a complex problem. Prior work has tackled individual subpieces of
the problem as highlighted by Figure~\ref{fig:Rel}. Monocle~\cite{perevsini2015monocle}, RuleScope~\cite{Rulescope}, and RuleChecker~\cite{zhang2017} use active probing
to verify whether the logical rules $R$\textsubscript{logical} 
are the same as the rules $R$\textsubscript{physical} of the data plane. 
ATPG~\cite{Zeng2014} creates test packets based on the control plane rules to verify whether paths taken by the packets on the data plane $P$\textsubscript{physical} are the
same as the expected path from the high-level policy $P$ without giving attention to the matched rules. 
VeriDP~\cite{Zhang2016} in SDNs and P4CONSIST~\cite{p4consist} in P4 SDNs use production traffic to only verify whether paths taken by the packets on the data plane $P$\textsubscript{physical} are the
same as the expected path from the control plane
$P$\textsubscript{logical}. NetSight~\cite{Handigol2014}, PathQuery~\cite{Narayana2016}, CherryPick~\cite{Tamanna2015}, and PathDump~\cite{pathdump} use production traffic whereas SDN Traceroute~\cite{Agarwal2014} uses active probes to verify $P \equiv P$\textsubscript{physical}.
Control plane solutions focus on verifying network-wide invariants such as reachability, forwarding loops, slicing, and
black hole detection against high-level network policies both for stateless and
stateful policies. This includes tools~\cite{fayaz2016buzz,Mai2011,Kazemian2012,Kazemian2013,Div,Canini2012,troubleshooting,veriflow,batfish,Lopes2015,Yang2016} that
monitor and verify some or all of the network-wide invariants by comparing the
high-level network policy with the logical rule set that translates to the logical path set at the control plane, i.e.,
$P \equiv R$\textsubscript{logical} or $P \equiv P$\textsubscript{logical}. These systems, however, only \textit{model} the
network behavior which is insufficient to capture firmware and hardware bugs as ``modeling'' and verifying the control-data plane consistency are significantly different techniques.

Typically, previous approaches to consistency checking proceed ``top-down,''
starting from what is
known to the management and control planes, and 
subsequently checking 
whether the data plane is consistent.
We claim that this is insufficient 
and underline this with several examples (\S\ref{sec:inc}) wherein data plane inconsistencies would go undetected. This can be critical, as analogous to security,
the overall system consistency is only as good as the weakest link in the chain.


We argue that we need to complement existing top-down approaches with a 
\emph{bottom-up} approach. To this end, we rely on an 
analogy to program testing. Programs
can have a huge state space, just like networks. There are
two basic approaches to test program correctness: 
one is static testing and the other is dynamic testing using \emph{fuzz testing} or fuzzing~\cite{Godefroid:2012:SWF:2090147.2094081}. Hereby,
the latter is often needed as the former cannot capture the actual run-time behavior. We realize that the same holds true for network state. 

Fuzz testing involves testing a program with invalid, unexpected, or random
data as inputs. 
The art of designing an effective fuzzer lies in generating 
semi-valid inputs that are \emph{valid enough} so that they are not directly 
rejected by the parser, but do create unexpected behaviors deeper in the program, 
and are \emph{invalid enough} to expose corner cases that have not 
been dealt with properly.
For a network, this corresponds to checking its behavior not only
with the expected production traffic but with \emph{unexpected or abnormal}
packets~\cite{shukla2}. However, in networking, what is expected or unexpected depends not only
on the input (ingress) port but also the topology till the exit (egress) port and configuration i.e., rules on the switches. Thus, there is a huge state space
to explore. Relying only on production traffic is not sufficient because production traffic may or may not trigger inconsistencies. However, having faults that can be triggered at any point
in time, due to a change in production traffic e.g., malicious or accidental, is
undesirable for a stable network. Thus, we need \emph{fuzz testing for checking
	network consistency}.  Accordingly, this paper introduces $\system$ which
combines such capabilities with previous approaches to verify SDNs 
against persistent data plane faults. Therefore, similar to program
testing we ask: \emph{``How far to go with
	consistency checks?''} 

\smartparagraph{Our Contributions:}\\
\vspace{-\topsep} 
\begin{itemize}[noitemsep,wide=0pt, leftmargin=\dimexpr\labelwidth + 2\labelsep\relax]
	\item We identify and categorize the causes and symptoms of data plane
	faults which are currently unaddressed to provide some useful insights into the limitations of existing approaches by investigating their frequency of occurence. Based on our insights, we make a case for fuzz testing mechanism for campus and private datacenter SDNs (\cref{sec:un});\\
	\item We introduce a novel methodology, \system which detects and later, localizes faults
	by comparing control vs.\ data plane information for all three
	components, rules, topology, and paths.
	It uses production traffic as well as active probes (to fuzz test
	the data plane state) (\cref{sec:Met});\\
	\item We develop and evaluate \system prototype\footnote{\system software and experiments with topologies and configs will be made public to the research community.} in multiple experimental topologies representative of multi-path/grid campus and private datacenter SDNs. Our results show that  
	fuzzing through \system outperforms baseline approach in all experimental topologies while consuming minimal resources as compared to Header Space Analysis to detect and localize data plane faults (\cref{sec:Imp});\\
\end{itemize}

\vspace{-1em}
\section{ Background \& Motivation}
\label{sec:un}
This section briefly navigates the landscape of faults and reviews the symptoms and causes (\cref{sec:1}) to set the stage for the program testing analogy in networks (\cref{sec:3}). Finally, we highlight the scenarios of data plane faults manifesting as inconsistency (\cref{sec:inc}).
\vspace{-1em}
\subsection{Landscape of Faults: Frequency, Causes and Symptoms}
\label{sec:1}

As per the survey~\cite{Survey}, the top primary causes for abnormal network behaviour or failures in the order of their frequency of occurence are the following:\\
\textit{1)} \textit{Software bugs}: code errors, bugs, etc.,
\textit{2)} \textit{Hardware failures or bugs}: bit errors or bitflips, switch failures, etc.,
\textit{3)} \textit{Attacks and external causes}: compromised security, DoS/DDos, etc., and
\textit{4)} \textit{Misconfigurations}: ACL /protocol misconfigs, etc.

In SDNs, the above causes still exist and are persistent~\cite{Zeng2014, bitflip, perevsini2015monocle, kuzniar2015you, cacheflow, jrex, soft, fib-ack, compromised,spook, blackhat}. We, however, realized that the symptoms~\cite{Survey} of the above causes can manifest either as functional or performance-based problems on the data plane. To clarify further, the symptoms are either functional (reachability, security policy correctness, forwarding loops, broadcast/multicast storms) or performance-based (Router CPU high utilization, congestion, latency/throughput, intermittent connectivity). To abstract the analysis, if we disregard the performance-based symptoms, we realize the functional problems can be reduced to the verification of network correctness. Making the situation worse, the faults manifest in the form of \emph{inconsistency} where the \emph{expected} network state at control plane is different to the \emph{actual} data plane state.

A physical network or network data plane comprises of devices and links. In SDNs, such devices are SDN switches connected through links. The data plane of the SDNs is all about the network behaviour when subjected to input in the form of traffic. Just like in programs, we need different test cases as inputs with different coverage to test the code coverage. Similarly, in order to \emph{dynamically} test the network behaviour thoroughly, we need input in the form of traffic with different \emph{coverage}~\cite{Varghese:2015:TPT:2838899.2823394}. Historically, the network correctness or functional verification on the data plane has been either a path-based verification (network-wide)~\cite{Zhang2016,Handigol2014,Narayana2016,Tamanna2015,pathdump,Agarwal2014,foces} or a rule-based verification (mostly switch-specific)~\cite{Zeng2014,perevsini2015monocle,Agarwal2014,zhang2017,Rulescope,sdnprobe}. A path-based verification can be end-to-end or hop-by-hop whereas rule-based verification is a switch-by-switch verification. The network coverage brings us to the concept of Packet Header Space Coverage. 
\begin{figure}[t]
	\vspace{-1em}
	\centering
	\includegraphics[width=.8\columnwidth]{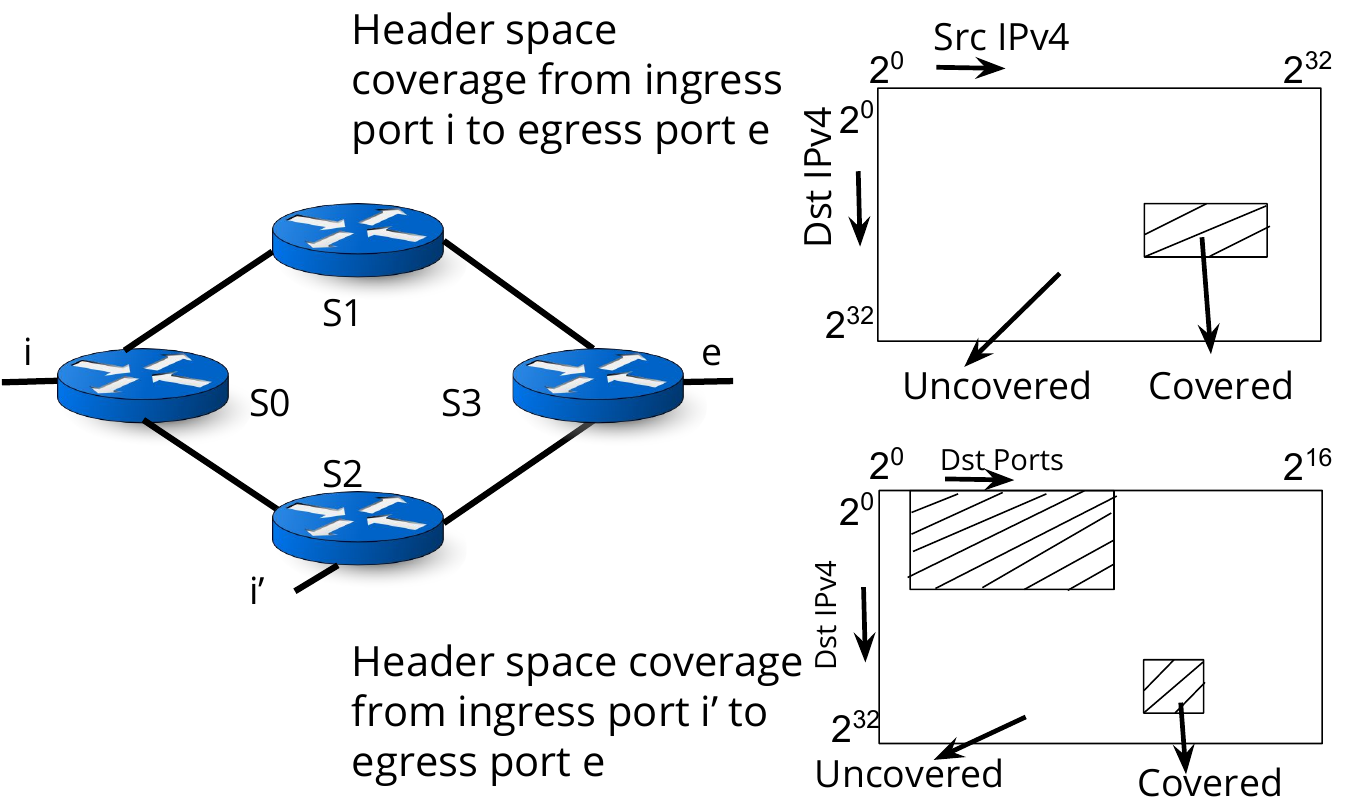}
	\caption{Example topology with two example ingress/egress port pairs (source-destination pairs)
		and their packet header space coverage (w.r.t IPv4 addresses and destination ports).}
	\label{fig:top}
\end{figure}

\subsection{Packet Header Space Coverage: Active vs Passive}
\label{sec:3}
\vspace{-.2em}
We observe that networks just as programs can have a huge distributed state space. Packets with
their packet headers, including source IP, destination IP, port numbers, etc.,
are the inputs and the state includes all forwarding equivalence classes defined
by the flow rules. Note, that every pair of ingress-egress ports (source-destination pair) can have
different forwarding equivalence classes (FECs). We use the term \emph{covered packet header space} to refer to it. 
Our motivation is that the forwarding
equivalence classes refer to parts of the packet header space. 
For a given pair of ingress and egress ports (source-destination pair),
when receiving traffic on the egress port from the ingress port, we can check if the
packet is covered by the corresponding ``packet header space''. If it is within
the space it is ``expected'', otherwise it is ``unexpected'' and, thus, we have
discovered an inconsistency due to the presence of a fault on the data plane.

Consider the example topology in Figure~\ref{fig:top}.  It consists of four
switches S0, S1, S2, and S3. Let us focus on two ingress ports $i$ and $i'$ and
one egress port $e$. The figure also includes possible packet header space coverage. For $i$ to $e$, it includes matches for the source and destination IPv4 addresses. For $i'$ to $e$, it includes matches for the destination IPv4 address and possible
destination port ranges. Indeed, it is possible to take into account other dimensions such as IPV6 header space, source-destination MAC addresses.

When testing a network, if traffic adheres to a specific packet header space, there are
multiple possible cases. If we observe a packet sent via an ingress port $i$ and received at an egress port $e$ then we need to check if it is within the covered area, if it is not we refer to the packet as ``unexpected'' and then, we have an inconsistency for that packet header space caused by a fault. If a packet from an ingress port is within the expected packet header space of multiple egress ports, we need to check if the sequence of rules \emph{expected to be matched} and path/s \emph{expected to be taken} by the packet correspond to the actual output port on data plane. This is yet another way of finding inconsistency caused by faults.
\begin{table}[t]
	\centering
	{\small
		\scalebox{0.75}{
			\begin{tabular}{ |p{2.3cm}|p{2cm}|p{1cm}|p{1cm}|}
				\hline
				& &  \multicolumn{2}{c|}{Type of monitoring/verification} \\ \cline{3-4}
				Related work in the data plane & Traffic Type (Packet header space coverage) & \multicolumn{1}{c|}{Rule-based} &  \multicolumn{1}{c|}{ Path-based}\\
				\hline
				ATPG~\cite{Zeng2014}~\tablefootnote{In this tool, if the packet is received at the expected destination from a source, path is considered to be the same.}  & Active &(\checkmark)& {(\checkmark)}\\
				\hline
				Monocle~\cite{perevsini2015monocle}   & Active& {(\checkmark)} & $\times$ \\
				\hline
				RuleScope~\cite{Rulescope} & Active & {(\checkmark)} & $\times$ \\
				\hline
				RuleChecker~\cite{zhang2017}\tablefootnote{\label{note2}In this tool, authors claim that tool may detect match and action faults \emph{without guarantee}.}  & Active & (\checkmark)& $\times$\\
				\hline
				SDNProbe~\cite{sdnprobe} & Active & {(\checkmark)} & {(\checkmark)} \\
				\hline
				FOCES~\cite{foces} & Passive &{(\checkmark)} & {(\checkmark)}\\
				\hline
				VeriDP~\cite{Zhang2016}& Passive  & {(\checkmark)} & {(\checkmark)}\\
				\hline
				NetSight~\cite{Handigol2014}  & Passive & {(\checkmark)} & {(\checkmark)}\\
				\hline
				PathQuery~\cite{Narayana2016} & Passive & {(\checkmark)} & {(\checkmark)}\\
				\hline
				CherryPick~\cite{Tamanna2015}\tablefootnote{In this tool, issues in only symmetrical topologies are addressed.}& Passive  & {(\checkmark)} & {(\checkmark)} \\
				\hline
				PathDump~\cite{pathdump}& Active & {(\checkmark)} & {(\checkmark)} \\
				\hline
			    P4CONSIST~\cite{p4consist} & Passive & $\times$ & {\checkmark} \\
				\hline
				\thead{\system} & Active, Passive & {\checkmark} & {\checkmark} \\
				
				\hline
		\end{tabular}}
		\caption{Classification of related work in the data plane based on the type of the verification and the packet header space coverage. \checkmark denotes full capability, (\checkmark) denotes a part of full capability, $\times$ denotes missing capability.}
		\label{tab:rel}	
		\vspace{-1em}
	}
\end{table}

Similar to program testing where negative test cases are used to fuzz test, in networking, we should not only test the network state with ``expected'' or the production traffic, but also with specially
crafted probe packets to test corner cases and ``fuzz test''. In principle, there are two ways for testing network forwarding: passive and active. Passive corresponds to using the existing traffic or production traffic while active refers
to sending specific probe traffic. The advantage of passive traffic is
that it has low overhead and popular forwarding paths are tested repeatedly.  However, production traffic may
(a)~not cover all cases (covers only faults that can be triggered by production traffic only); (b)~change rapidly; and (c)~have delayed fault detection, as the fraction of traffic triggering the faults is delayed. Indeed, malicious users may be able to inject malformed traffic that may trigger fault/s. Thus, production traffic may not cover the whole packet header space achievable by active probing.

Furthermore, we should also fuzz test the network state. This is important as we
derive our network state from the information of the controller. Yet,
this is not sufficient since we cannot presume that the controller state is
complete and/or accurate. Thus, we propose to generate packets that are
outside of the covered packet header space of an ingress/egress port pair. We
suggest doing this by systematically and continuously or periodically testing the header space just outside of
the covered header space. E.g., if port 80 is within the covered header space
test for port 81 and 79. If \texttt{x.0/17} is in the covered header space test for
\texttt{x.1.0.0} which is part of the \texttt{x.1/17} prefix. In addition, we propose to randomly
probe the remaining packet header space continuously or periodically by generating appropriate test traffic.
The goal of active traffic generation through fuzzing is to detect the faults identifiable by active traffic only.

Table~\ref{tab:rel} shows the existing data plane approaches on the basis of kind of verification or monitoring in addition to the packet header space coverage. We see that the existing dataplane verification approaches are insufficient when it comes to both path and rule-based verification in addition to ensuring sufficient packet header space coverage. In this paper, our system \system aims to ensure packet header space coverage in addition to path and rule-based verification to ensure network correctness on the data plane and thus, detecting and localizing persistent inconsistency.
\begin{figure}[t]
	\vspace{-1em}
	\centering
	\includegraphics[width=0.8\columnwidth]{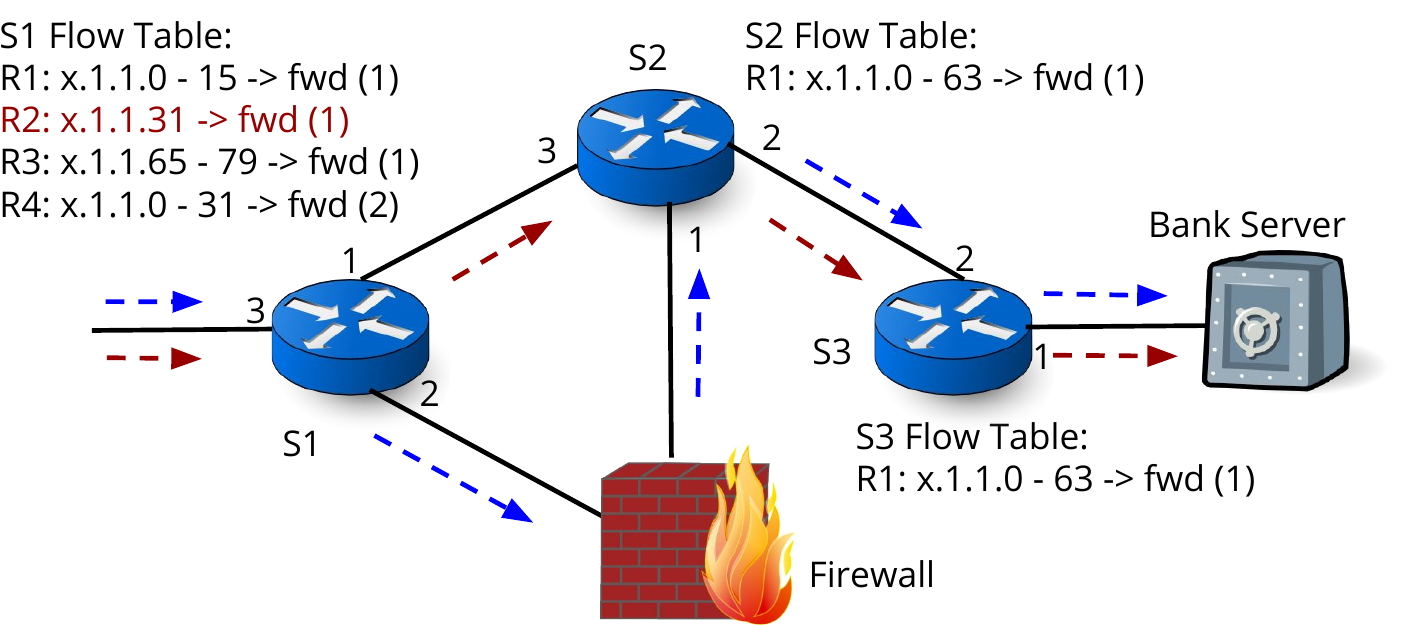}
	\caption{Example misconfiguration with a hidden rule. \\Expected/actual route
		$\Leftrightarrow$ blue/red arrows.}
	\label{fig:space}
\end{figure}
\vspace{-1em}
\subsection{Data plane faults manifesting as inconsistency}
\label{sec:inc}
\subsubsection{Faults identified by Passive Traffic: Type-p}
\label{sec:act}
To highlight the type of faults, consider a scenario shown in Figure~\ref{fig:space}. It has three
OpenFlow switches (S1, S2, and S3) and one firewall (FW).  Initially, S1 has
three rules R1, R3, and R4. R4 is the least specific rule and has the lowest
priority. R1 has the highest priority. Note the rules are written in the order of their priority.
\vspace{-.01em}

\smartparagraph{Incorrect packet trajectory:}
We start by considering a \emph{known}
fault~\cite{Zhang2016, Handigol2014, pathdump}--\textbf{hidden rule/misconfiguration}.
For this, the rule R2 is added to S1 via the switch command line utility. The
controller will remain unaware of R2 since R2 is a non-overlapping flow
rule. Thus, it is installed without notification to the
controller~\cite{onefive}. \cite{perevsini2015monocle, kuzniar2015you} have hinted at this problem. As a result, traffic to IP \texttt{x.1.1.31} bypasses the firewall as it uses a different path.

\textbf{Priority faults} are another reason for such incorrect forwarding where either rule priorities get swapped or are not taken into account. The Pronto-Pica8 3290 switch with PicOS 2.1.3 caches rules without accounting for rule priorities~\cite{cacheflow}. The HP ProCurve switch lacks rule priority support~\cite{kuzniar2015you}. Furthermore, priority faults may manifest in many forms e.g., they may cause the trajectory changes or incorrect matches even when the trajectory remains the same. \textbf{Action faults} can be another reason where bitflip in the action part of the flowrule may result in a different trajectory.

\emph{\textbf{Insight 1:} Typically, the packet trajectory tools only monitor the path.}
\vspace{-.5em}

\smartparagraph{Correct packet trajectory, incorrect rule matching:}
If we add a higher priority rule in a similar fashion where the path
does not change, i.e., the match and action remains the same as in the shadowed
rule, then previous work will be unable to detect it and, thus, it is
\emph{unaddressed}~\footnote{We validated this via experiments. OpenFlow
	specification~\cite{onefive} states that if a non-overlapping rule
	is added via the switch command-line utility, controller is not notified.}. Even if the packet trajectory is correct but wrong rule is matched, it can inflict serious damages. Misconfigs, hidden rules, priority faults, match faults (described next) may be the reason for incorrect matches.
Next, we focus on \textbf{match faults} where anomaly in the match part of a forwarding flow rule on a switch causes the packets to be matched incorrectly. We again highlight \emph{known} as well as \emph{unaddressed} cases starting with a known scenario. In Figure~\ref{fig:space}, if a bitflip\footnote{A previously unknown firmware bug in HP 5406zl switch.}, e.g., due to hardware problems, changes R1 from \texttt{x.1.1.0/28} to match from \texttt{x.1.1.0} upto \texttt{x.1.1.79}.  Traffic to \texttt{x.1.1.17} is now forwarded based on R1 rather than R4 and thus, bypasses the firewall. This may still be detectable, e.g., by observing the path of a test
packet~\cite{Zhang2016}.  However, the bitflip in R1
also causes an overlap in the match of R1 and R3 in switch S1 and both rules have the
same action, i.e., forward to port 1.  Thus, traffic to \texttt{x.1.1.66} supposed to be matched by
R3 will be matched by R1. If later, the network administrator removes R3, the traffic still pertaining to R3 still runs. This violates the network policy. In this paper, we categorize the dataplane faults detectable by the production traffic as \emph{Type-p faults}.

\emph{\textbf{Insight 2:} Even if the packet trajectory remains the same, the matched rules need to be monitored. }
\begin{figure}[t!]
	\vspace{-1em}
	\centering
	\includegraphics[width=0.8\columnwidth]{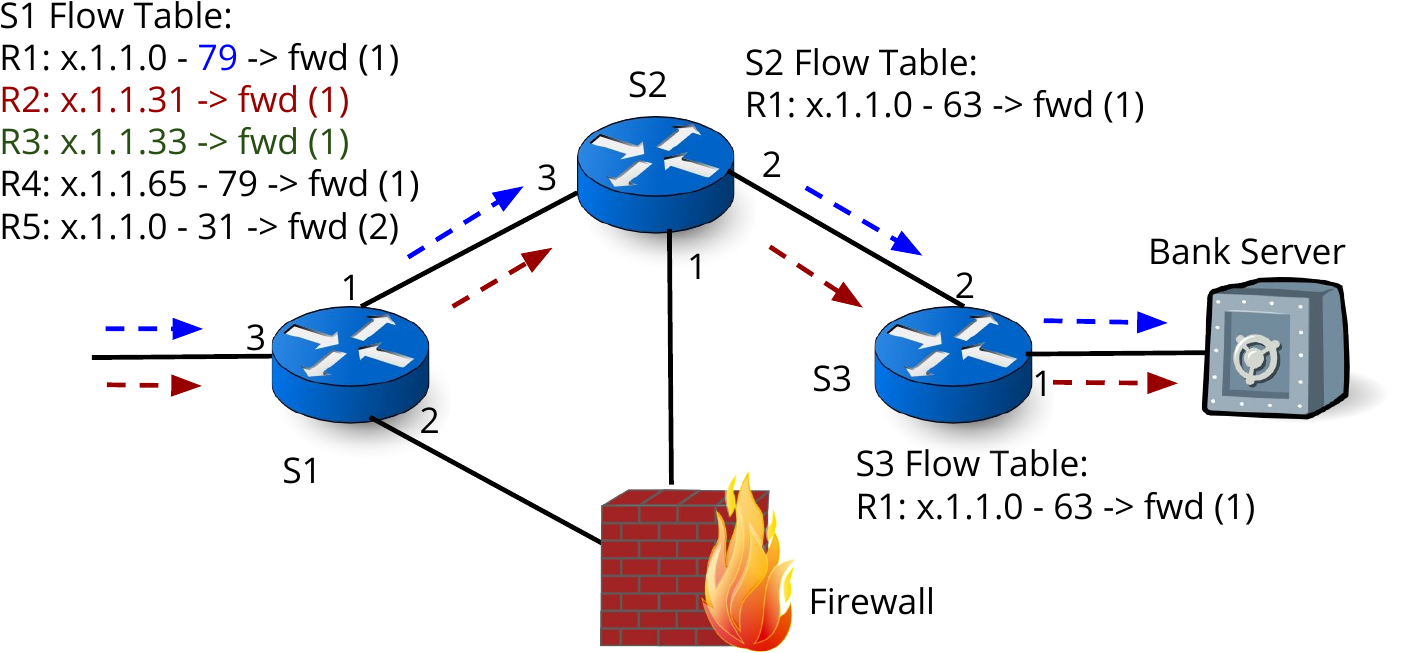}
	\caption{Example misconfiguration with a hidden rule detectable by active probing only. Expected/actual route
		$\Leftrightarrow$ blue/red arrows.}
	\label{fig:dia3}
\end{figure}
\subsubsection{Faults identified by Active Traffic only: Type-a}
\label{sec:2}
To highlight, we focus on hidden or misconfigured rule R3 (in green) in Figure~\ref{fig:dia3}. This rule matches the traffic corresponding to \texttt{x.1.1.33} on switch S1 and reaches the confidential bank server, however, the expected traffic or the production traffic does not belong to this packet header space~\cite{compromised,spook, blackhat}. Therefore, we need to generate probe packets to trigger such rules and thus, detect their presence. This will require generating and sending the traffic corresponding to the packet header space which is not expected by the control plane. We call this traffic as \emph{fuzz traffic} in the rest of the paper since it tests the network behavior with unexpected packet header space. In this paper, we categorize the dataplane faults detectable by only the active or fuzz traffic as \emph{Type-a faults}.

\emph{\textbf{Insight 3:} The tools which test the rules check only rules ``known'' to the control plane (SDN controller) by generating active traffic for ``known'' flows.}

\emph{\textbf{Insight 4:} Typically, the active traffic for certain flows checks only if the path remains the same even when rule/s matched may be different on the data plane.}

\vspace{-.75em}
\section{\system Methodology}
\label{sec:Met}
Motivated by our insights gained in \cref{sec:inc} about the Type-p and Type-a faults on the data plane resulting in inconsistency, we aim to take the consistency checks further. Towards this end, our novel methodology, \system compares forwarding rules, topology, and paths of the control and the data plane, using top-down and bottom-up approaches, to detect and localize the data plane faults. 

\begin{figure}[!tbp]
	\centering
	\includegraphics[width=0.9\columnwidth]{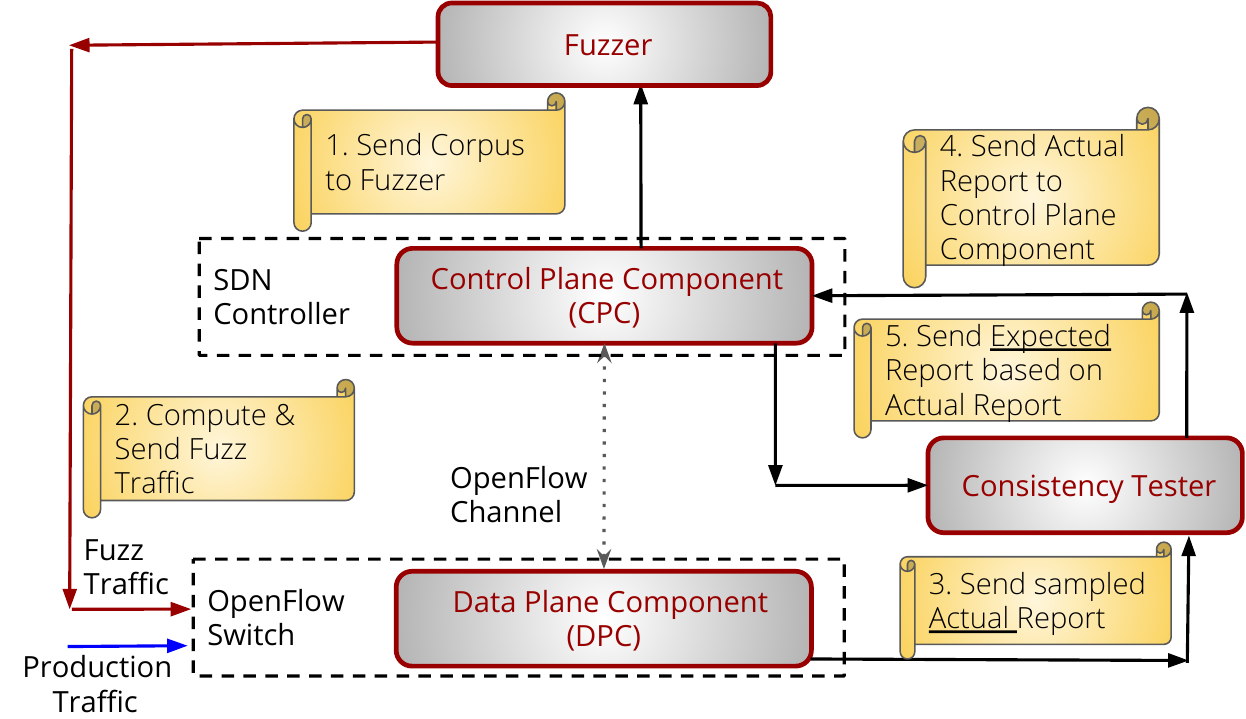}
	\caption{\system Methodology.}
	\label{fig:sys}
	\vspace{-.5em}
\end{figure} 

$\system$, derived from \textit{P}assive and \textit{A}ctive (fu\textit{ZZ} testing), 
takes into account both production and probe traffic to ensure adequate packet header 
space coverage.
\system checks the matched forwarding flow rules as well as the links constituting paths of various packet headers (5-tuple microflow) present in the passive and active traffic. 
To detect faults, \system 
collects state information (in terms of \emph{reports})
from the control and the data plane: \system compares 
the ``expected'' state reported by 
the control to the ``actual'' state collected from the data plane. Figure~\ref{fig:sys} illustrates the \system methodology. It consists of four components sequentially:\\
\begin{enumerate}[noitemsep,wide=0pt, leftmargin=\dimexpr\labelwidth + 2\labelsep\relax]
	\vspace{-1em}
	\item \textbf{Control Plane Component (\cp):} Uses the current controller information
	to proactively compute the packets that are reachable between any/every source-destination pair. It then sends the corpus of seed inputs to Fuzzer. For any given packet header and source-destination pair, it reactively generates an expected report which encodes the paths and sequence of rules. (\cref{sec:CPC})\\
	\item \textbf{\fuzz:} Uses the information from \cp to compute the packet header space not covered by the controller and hence, the production traffic. It generates active traffic for fuzz testing the network. (\cref{sec:fuzz})\\
	\item \textbf{Data Plane Component (\dpl):} For any given packet header and source-destination pair, it encodes the path and sequence of forwarding rules to generate a \emph{sampled} actual report. (\cref{sec:DPC})\\
	\item \textbf{\consistencychecker:} Detects and later, localizes
	faults by comparing the expected report/s from the \cp with the actual report/s from the \dpl. (\cref{sec:cc})\\ 
\end{enumerate}
\vspace{-1em}
Now, we will go through all components in a non-sequential manner for the ease of description.
\vspace{-.5em}
\subsection{Data Plane Component (DPC)} 
\label{sec:DPC}
To record the actual path of a packet and the rules that are matched in forwarding, we rely on tagging the packets contained in active and production traffic.  
In particular, we propose the use of a shim header that 
gives us sufficient space even for larger network diameters or
larger flow rule sets. Indeed, INT~\cite{kim2015band} can be used for data plane monitoring, however, it is applicable for P4 switches~\cite{Bosshart2014} only. Unlike~\cite{Narayana2016, Zhang2016, Tamanna2015,pathdump,Handigol2014}, we use our custom shim header for tagging, therefore tagging is possible without limiting forwarding capabilities. To avoid adding extra monitoring rules on the scarce TCAM which may also affect the forwarding behavior~\cite{Narayana2016, Zhang2014}, we augment OpenFlow with new actions. 
Between any source-destination pair, the new actions are used by all rules of the switches to add/update the shim header if necessary for encoding the sequence of inports (path) and matched rules. To remove the shim header, we use another custom OpenFlow action. To trigger the actual report to the \consistencytester, we use \emph{sFlow}~\cite{sflow} sampling. Indeed, we can use any other sampling tool. Note sFlow is a \emph{packet sampling} technique so it samples packets not flows based on sampling rate and polling interval. For a given source, the report contains the packet header, the shim header content, and the egress port of the exit switch (destination).

Even with a shim header: \prot, however, it is 
impractical to expect packets to have available and sufficient space 
to encode information about each port and rule on the path. 
Therefore, we rely on a combination of bloom filter~\cite{bloom2} and binary hash chains. For scalability purposes, sampling is used before sending a report to the
\consistencytester. 
\begin{algorithm}[t]
	\scriptsize
	\caption{Data Plane Tagging}
	\label{alg:Tag}
	\SetKwInOut{Input}{Input}\SetKwInOut{Output}{Output}
	\Input{($p$,
		$s$, $i$, $o$, $r$) for each incoming packet $p$ and switch
		with ID $s$ let $i$ be the inport ID and $o$ the outport ID
		for packet $p$, $r$ is the flow rule used for forwarding.}
	\Output{Tagged packet $p$ if necessary with the \prot shim header.}
	
	\tcp{Is there already a shim header, e.g., ($s$,$i$) is not an entry
		point or source port} 
	\If{($p$ has no shim header)} {
		\tcp{Add shim header with ``Ethertype'' $2080$,
			initialize tag values- \vp: entry point hash, \vr: 1.}
		$p.push\_verify$;}
	\tcp{Determine \up ID from switch ID $s$ and port ID $i$}
	\up = $s \parallel i$;
	
	\tcp{Bloom Filter} 
	$p.\vp \leftarrow bloom(hash$(\up))$;$
	
	\tcp{Determine \ur ID from rule ID $r$ of table ID $t$} 
	\ur = $s \parallel r \parallel t$;
	
	\tcp{Binary hash chain} 
	$p.\vr \leftarrow hash$($p.\vr$, \ur)$;$
	
	\tcp{Shim header has to be removed if ($s,o$) is
		exit point}
	\If{(($s,o$) is exit point)} {
		\If{($p$ has no shim header)} {
			\tcp{For traffic injected between a source-destination pair}
			$p.push\_verify$;}
		$Generate\_report((s,o), p.\vp, p.\vr,\newline p.header$)$;$
		$p.pop\_verify$;
	}
\end{algorithm}
\vspace{-0.4em}

\begin{figure}[t]
	\vspace{-.5em} 
	\centering
	\includegraphics[width=.8\columnwidth]{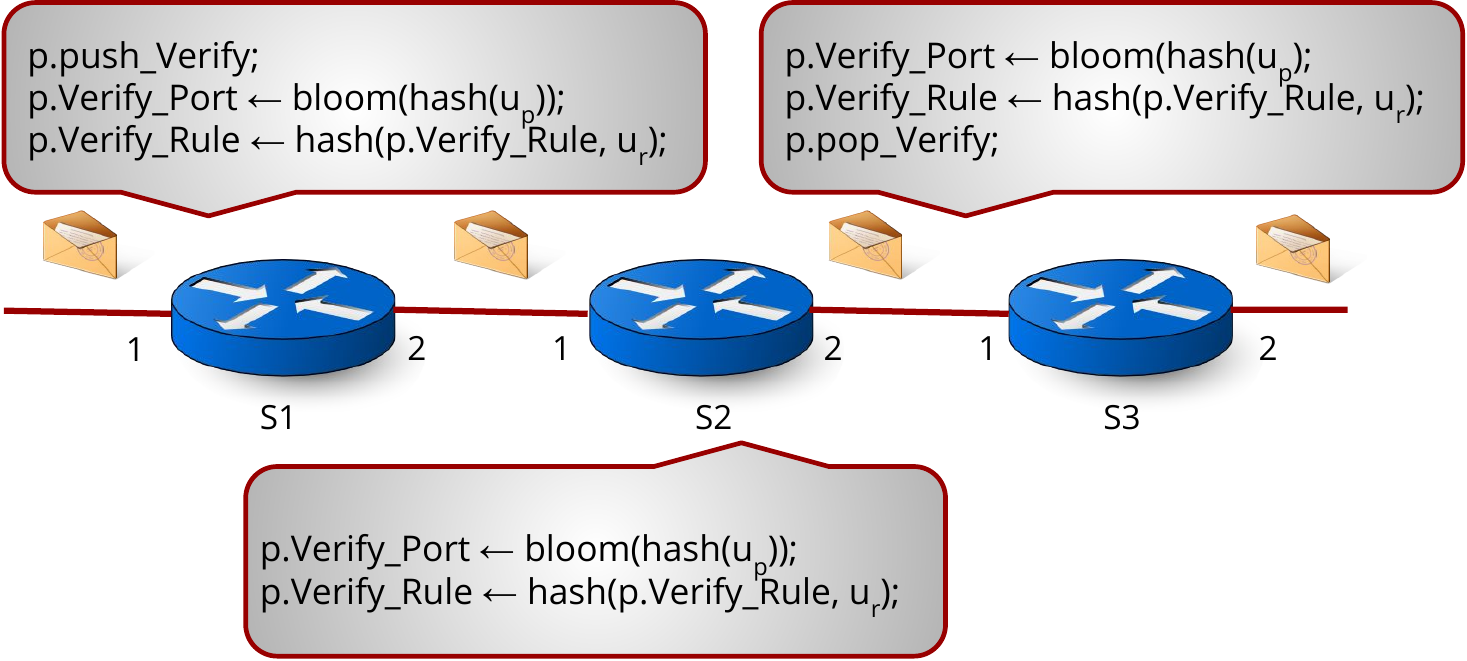}
	\caption{Data plane tagging using bloom filters and hashing.}
	\label{fig:tag}
	\vspace{-.5em} 
\end{figure}

\smartparagraph{Data Plane Tagging:}  To limit the overhead, we decided to insert \prot shim header on layer-2. \vt is EtherType for \prot header, \vpd (bloom filter) encodes the local inport in a switch, and \vrd (binary hash chain) encodes the local rule/s in a switch. Thus, the encoding is done with the help of the bloom filter and binary hash chain respectively. To take actions on the proposed \prot shim header and to save TCAM space, we propose four new OpenFlow actions: two for adding (\textit{push\_verify}) and removing (\textit{pop\_verify}) the \prot shim header and two for updating the \vpd (\textit{set\_\vp}) and \vrd (\textit{set\_\vp}) header fields respectively.  Since, the header size of the \vpd and \vrd and tagging actions are implementation-specific, we have explained them in prototype section \cref{sec:DPCI}, \cref{sec:actions} respectively. Algorithm~\ref{alg:Tag} explains the data plane tagging algorithm between a source-destination pair. For each packet either from the production or active traffic (\cref{sec:fuzz}) entering the source inport, \prot shim header will be added automatically by the switch. For each switch on the path, the tags in the packet namely, \vpd and \vrd fields get updated automatically. 
Figure~\ref{fig:tag} illustrates the per-switch tagging approach. Once the packet leaves the destination outport, the resulting report known as the \emph{actual report} is sent to the \consistencytester (\cref{sec:cc}). Note if there is no \prot header, \prot shim header is pushed on the exit switch to ensure that any traffic injected at any switch interface between a source-destination pair gets tagged. To reduce the overhead on the \consistencytester as well as on the switch, we employ sampling at the egress port. 
We continuously or periodically test the network as the data plane is dynamic due to reconfigurations, link/switch/interface failures, and topology changes. Note, periodic testing can also be executed by implementing timers.
\vspace{-.75em}
\subsection{Control Plane Component (CPC)}
\label{sec:CPC}
In principle, we can use the existing control plane mechanisms, including HSA~\cite{Kazemian2012}, NetPlumber~\cite{Kazemian2013} and APVerifier~\cite{Yang2016}. In addition to experiments in~\cite{Yang2016}, our independent experiments show that Binary Decision Diagram (BDD)-based~\cite{Bryant:1986:GAB:6432.6433} solutions like~\cite{Yang2016} perform better for set operations on headers than HSA~\cite{Kazemian2012} and NetPlumber~\cite{Kazemian2013}. In particular, we will propose in the following a novel BDD-based solution that supports rule verification in addition to path verification (APVerifier~\cite{Yang2016} takes into account only paths).
Specifically, our Control Plane Component (CPC) performs two functions: a) \emph{Proactive} reachability and corpus computation, and b) \emph{Reactive} tag computation. 

\smartparagraph{\emph{Proactive} Reachability \& Corpus Computation:} We start by introducing an abstraction of a single switch configuration called \emph{switch predicate}. 
In a nutshell, a switch predicate specifies the forwarding behavior of the switch for a given set of incoming packets, and is defined in turn by the \emph{rule predicates}.
More formally, the general configuration abstraction of a SDN switch $s$ 
with ports $1$ to $n$ can be described by switch predicates: $S_{\scriptscriptstyle i,j}$ where $i \in \{1, 2, . . . , n\}$ and $j \in \{1, 2, . . . , n\}$ where $n$ denotes the number of switch ports. The packets headers satisfying predicate $S_{\scriptscriptstyle i,j}$ can be forwarded from port $i$ to port $j$ only. 
The switch predicate is defined via rule predicates: $R_{\scriptscriptstyle i,j}$ which are given by the flowrules belonging to the switch $s$ and a flowtable $t$. 
Each rule has an identifier that consists of a $unique\_id$ and $table\_id$ representing the flowtable $t$ in which the rule resides, $in\_por$t array representing a list of inports for that rule, $out\_port$ array representing a list of outports in the action of that rule and the rule priority $p$. Based on the rule priority $p$, $in\_port$ in the match part and $out\_port$ in the action part of a flowrule, each rule has a list of rule predicates (BDD predicates) which represent the set of packets that can be matched by the rule for the corresponding inport and forwarded to the corresponding outport. 

Similar to the plumbing graph of~\cite{Kazemian2013}, we generate a dependency graph of rules (henceforth, called rule nodes) called \emph{reachability graph} based on the topology and switch configuration which computes the set of packet headers between any source-destination pair. There exists an edge between the two rules $a$ and $b$, if (1) out\_port of rule $a$ is connected to in\_port of $b$; and (2) the intersection of rule predicates for $a$ and $b$ is non-empty. For computational efficiency, each rule node keeps track of higher priority rules in the same table in the switch. A rule node computes the match of each higher priority rule, subtracting it from its own match. 
We refer to this as the \emph{same-table dependency} of rules in a switch. In the following, by slightly abusing the notation, we will use switch predicates $S_{\scriptscriptstyle i,j}$ and rule predicates $R_{\scriptscriptstyle i,j}$ to denote also the set of packet headers: $\{p_1, p_2,...,p_n\}$ they imply.
Disregarding the ACL predicates for simplicity, the rule predicates in each switch $s$ representing packet header space forwarded from inport $i$ to outport $j$ is given by $R^{\scriptscriptstyle fwd}_{\scriptscriptstyle i,j}$. The switch predicates are then computed as: $S_{\scriptscriptstyle i,j} = \cup_{\scriptscriptstyle i,j} R^{\scriptscriptstyle fwd}_{\scriptscriptstyle i,j}$


More specifically, to know the reachable packet header space (set of packet headers) between any source-destination pair in the network, we inject a fully-wildcarded packet header set \emph{h} from the source port. If the intersection of the switch predicate $S_{\scriptscriptstyle i,j}$ and the current packet header $p$ is non-empty, i.e., $S_{\scriptscriptstyle i,j} \cap \{p\} \neq \phi$, the packet is forwarded to the next switch until we reach the destination port. Thus, we can compute reachability between any/every source-destination pair. For caching and tag computation, we generate the inverse reachability graph simultaneously to cache the traversed paths and rules matched by a packet header $p$ between every source-destination pair. After the reachability/inverse reachability graph computation, \cp sends the current switch predicates of the entry and exit switch pertaining to a source-destination pair to \fuzz as a corpus for fuzz traffic generation (\cref{sec:fuzz}). 

In case of a FlowMod, the reachability/inverse reachability graph and new corpus are re-computed. Recall every rule node in a reachability graph keeps track of high-priority rules in a table in a switch. Therefore, only a part of the affected reachability/inverse reachability graph needs to be updated in the event of rule addition/deletion. In the case of rule addition, the same-table dependency of the rule is computed by comparing the priorities of new and old rule/s before it is added as a new node in the reachability graph. If the priority of a new rule is higher than any rule/s and there is an overlap in the match part, the new switch predicate: $S'_{\scriptscriptstyle i,j}$ as per the new rule predicate: $R'^{\scriptscriptstyle fwd}_{\scriptscriptstyle i,j}$ is computed as:\\
$S'_{\scriptscriptstyle i,j} =  R'^{\scriptscriptstyle fwd}_{\scriptscriptstyle i,j} \cup (R^{\scriptscriptstyle fwd}_{\scriptscriptstyle i,j} - R'^{\scriptscriptstyle fwd}_{\scriptscriptstyle i,j})$


\smartparagraph{\emph{Reactive} Tag Computation:} For any given data plane report corresponding to a packet header $p$ between any source-destination pair, we traverse the pre-computed inverse reachability graph to generate a list of sequences of rules that can match and forward the actual packet header observed at a destination port from a source port. Note, there can be multiple possible paths, e.g., due to multiple entry points and per-packet or per-flow load balancing. For a packet header $p$, the appropriate \vpd and \vrd tags are computed similarly as in Algorithm~\ref{alg:Tag}. The expected report is then sent to the \consistencychecker (\cref{sec:cc}) for comparison. Note we can generate expected reports for any number of source-destination pairs.
\vspace{-1em}
\subsection{\fuzz}
\label{sec:fuzz}
Inspired by the code coverage-guided fuzzers like LibFuzz~\cite{lib}, we design a mutation-based fuzz testing~\cite{mut} component called \fuzz. \fuzz receives the corpus of seed inputs in the form of the switch predicates of the entry and exit switch from the \cp for a source-destination pair. In particular, the switch predicates pertaining to the inport of the entry switch (source) and the outport of the exit switch (destination) represent \emph{expected} covered packet header space containing the set of packet headers satisfying those predicates. \fuzz applies mutations to the corpus (Algorithm~\ref{alg:Fuzz}). 

\smartparagraph{Where Can Most Faults Hide:} Before explaining Algorithm~\ref{alg:Fuzz}, we present a scenario to explain the packet header space area where potential faults can be present. Consider the example topology illustrated in Figure~\ref{fig:top}. Due to a huge header space in IPv6 (128-bit), we decide to focus on the destination IPv4 header space (32-bit) in a case of destination-based routing. In principle, it is possible to consider IPv6 header space as the fuzzing mechanism remains the same. We use $S_i$, $S_1$ and $S_e$ to represent covered packet header space (switch predicates) of switches $S0$, $S1$ and $S3$ between \emph{i-e} (source-destination pair). Note there can be multipaths for the same packet header \emph{p}. Now, assume there is only a single path: $S0 \rightarrow S1 \rightarrow S3$, the reachable packet header space or net covered packet header space area is given by $S_i \cap S_1 \cap S_e$. Note this area corresponds to the control plane perspective so there may be more or less coverage on the data plane. The production traffic is generated in the area $S_i$ which depends on the \emph{expected} rules of $S0$ at an ingress port \emph{i} for a packet header \emph{p} destined to \emph{e}. In principle, the production traffic will cover the packet header space area $S_i$. Now, the active traffic should be generated for the uncovered area $U - S_i (entry switch)$ where $U$ represents the universe of all possible packet header space which is $2^0$ to $2^{32}$ for a destination IPv4 header space.
\begin{algorithm}[t]
	\scriptsize
	\caption{\fuzz}
	\label{alg:Fuzz}
	\SetKwInOut{Input}{Input}\SetKwInOut{Output}{Output}
	\Input{Switch predicates of entry ($S_i$) and exit $S_e$ switch for a source-destination pair $i$-$e$}
	\Output{$fuzz\_sweep\_area, random\_fuzz\_area$}
	\tcp{Generate fuzz traffic in the difference of covered packet header space area between entry and exit switch}
	$fuzz\_sweep\_area \leftarrow S_e - S_i$;\\
	$residual\_area \leftarrow U - fuzz\_sweep\_area - S_i$;\\
	$random\_fuzz\_area \leftarrow \Phi$;\\
	\tcp{Generate fuzz traffic in completely uncovered packet header space area randomly}
	\While {$random\_fuzz\_area \neq residual\_area$}
	{\do $fuzz \leftarrow random\_choose(residual\_area)$;\\
		$random\_fuzz\_area \leftarrow random\_fuzz\_area \bigcup fuzz$;\\
	}
\end{algorithm}
As stated in~\cref{sec:3}, we need to start with active traffic generation on the boundary of the net covered packet header space area between a source-destination pair as there is a maximum possibility of faults in this area. A packet will reach $e$ from $i$ iff all of the rules on the switches in a path match it; else it will be dropped either midway or on the first switch. Therefore, for an end-to-end reachability, the ruleset on $S0$ and $S3$ should match the packet $p$ contained in the production traffic belonging to the covered packet header space: $S_i \cap S_e$. This implies that we need to first generate active traffic in the area: $S_e - S_i$ and then randomly generate in the leftover area. Traffic can, however, be also injected at any switch on any path between a source-destination pair, thus the checking needs to be done for different source-destination pairs. 

We now explain the Algorithm~\ref{alg:Fuzz} in the context of Figure~\ref{fig:top}. For active or fuzz traffic generation, if there is a difference in the covered packet header space areas of S0 and S3, we first generate traffic in the area, i.e., $S_e - S_i$ denoted by $fuzz\_sweep\_area$ (Line 1). Recall there is a high probability that there may be hidden rules in this area since the header space coverage of the exit switch may be bigger than the same at the entry switch. Later, we generate traffic randomly in the residual packet header space area, i.e., $U - fuzz\_sweep\_area - S_i$ denoted by $residual\_area$ (Lines 2-6). We generate traffic randomly in the area as this is mostly, a huge space and fault/s can lie anywhere. The fuzz traffic generated randomly is given by the completely uncovered packet header space area denoted by $random\_fuzz\_area$. Thus, the fuzz traffic that the \fuzz generates belongs to the packet header space area given by $fuzz\_sweep\_area$ and $random\_fuzz\_area$. It is worth noting that not all of the packets generated by the fuzz traffic are allowed in the network due to a default drop rule in the switches. Therefore, if some packets in the fuzz traffic are matched, the reason can be attributed to either the presence of faulty rule/s, wildcarded rules or hardware/software faults to match such traffic. This also highlights that the fuzz traffic may not cause network congestion. As discussed previously, there is another scenario where the traffic gets injected at one of the switches on the path between a source-destination pair and may end up getting matched in the data plane. Verify header is pushed at the exit switch if it is not already present (\cref{sec:DPC}, \cref{sec:actions}) and thus, the packets still get tagged in the data plane to be sent in the actual report. However, the \cp may generate empty \vrd and \vpd tags as the traffic is unexpected. In such cases, the fault is still detected but may not be localized \emph{automatically} (\cref{sec:cc}). Furthermore, \fuzz can be positioned to generate traffic at different inports to detect more faults in the network between \emph{any/every} source-destination pair. In case if the production traffic does not cover all of the expected rules at the ingress or entry switch, \fuzz design can be easily tuned to also generate the traffic for critical flows. Our evaluations confirm that an exhaustive active traffic generator which randomly generates the traffic in the uncovered area performs poorly against \system in the real world topologies (\cref{sec:I2}). Note if the network topology or configuration changes, the \cp sends the new corpus to the \fuzz and Algorithm~\ref{alg:Fuzz} is repeated. We can continuously or periodically test the network with fuzz traffic for any changes.
\vspace{-.75em}
\subsection{\Consistencychecker}
\label{sec:cc}
\begin{algorithm}[t]
	\scriptsize
	\caption{\consistencychecker (\emph{detection, localization})}
	\label{alg:loc}
	\SetKwInOut{Input}{Input}\SetKwInOut{Output}{Output}
	\Input{Actual and expected report containing the $\vr_a$ and
		$\vp_a$ tags for a packet $p$ pertaining to a flow (5-tuple)}
	\Output{Detected and localized faulty switch $S_f$ or Faulty Rule $R_f$.}
	
	\tcp{Different rules were matched on data plane.} 
	\If{($\vr_a \neq \vr_e$)} {
		\tcp{Fault is detected and reported.}
		Report fault\\
		\tcp{Different path was taken on data plane.}
		\If{($\vp_a \neq \vp_e$)}{
			\tcp{Localize the fault.}
			\For{$i\gets0$ \KwTo $n$ \KwBy $1$}{
				\If{$\vp_a \cap \vp_{e^i} = \vp_{e^i}$}{No problems for this switch hop}
				\tcp{Previous switch wrongly routed the packet.}
				\Else{$S_f \leftarrow S_{i-1}$}}
		}
		\tcp{Path is same even rules matched are different.}
		\Else {
			\tcp{Localize Type-p match fault.}
			\If{$\vr_e \neq 0$}{Go through the different switches hop-by-hop to find $R_f$}
			\tcp{Localize Type-a fault.}
			\Else{ $R_f$ lies in $S0$ else go through the different switches hop-by-hop}
		}
	}
	\tcp{Different path was taken on data plane.}
	\ElseIf{($\vp_a \neq \vp_e$)}{
		\tcp{Type-p action fault is detected and reported.}
		Report fault\\
		\tcp{Localize Type-p action fault.}
		\For{$i\gets0$ \KwTo $n$ \KwBy $1$}{
			\If{$\vp_a \cap \vp_{e^i} = \vp_{e^i}$}{No problems for this switch hop}
			\tcp{Previous switch wrongly routed the packet.}
			\Else{$S_f \leftarrow S_{i-1}$}}
	}
	\Else{No fault detected}
\end{algorithm}
After receiving an actual report from the data plane, 
the \consistencychecker queries the \cp for its expected report for the packet header and the corresponding source-destination pair in the actual report. Once \consistencytester has received both reports, it compares both reports as per Algorithm~\ref{alg:loc} for fault detection and the localization. To avoid confusion, we use $\vp_a$, $\vr_a$ tags for the actual data plane report and $\vp_e$, $\vr_e$ tags for the corresponding expected control plane report respectively. If \vrd tag is different for a packet header and a pair of ingress and egress ports, then the fault is detected and reported (Lines 1-2). Note that we avoid the bloom filter false positive problem by first matching the hash value for the \vrd tag. Therefore, the detection accuracy is high unless a hash collision occurs in \vrd field (\cref{sec:hash}). Once a fault is detected, \ct uses the \vpd bloom-filter for localization of faults where the actual path is different from the expected path, i.e., the $\vp_a$ bloom filter is different from the $\vp_e$ bloom filter (Lines 3-8). Therefore, $\vp_a$ is compared with the per-switch hop \vpd in the control plane or $\vp_{e^i}$ for the $i\textsuperscript{th}$ hop starting from the source inport to the destination outport. This hop-by-hop walkthrough is done by traversing the reachability graph at the \cp hop-by-hop from the source port of the entry switch to the destination port of the exit switch. As per the Algorithm~\ref{alg:loc}, the bitwise logical AND operation between the $\vp_a$ and the $\vp_{e^i}$ is executed at every hop. 
It is, however, important to note that if actual path was same as expected path, i.e., $\vp_e =$ $ \vp_a$ even when actual rules matched were different on the data plane, i.e., $\vp_e \neq \vp_a$ (Lines 10-13), the localization of faulty $R_f$ gets tricky as it can be either a case of Type-p match faults (e.g., bitflip in match part) (Lines 10-11) or Type-a fault (Lines 12-13). Hereby, it is worth noting that there will be no expected report from the \cp in the case of \emph{unexpected} fuzz traffic. Therefore, \ct checks if $\vr_e \neq 0$ (Line 10). If true, localization can be done through hop-by-hop manual inspection of \emph{expected} switches or manual polling of \emph{expected} switches (Lines 10-11) else the Type-a fault \emph{may} be localized to the entry switch as it has a faulty rule that allows the unexpected fuzz traffic in the network (Lines 12-13). There is another scenario where the actual rules matched are same as expected but the path is different (Lines 14-20). This is a case of Type1-action fault (e.g., bitflip in the action part of the rule). In this case, the expected and actual \vpd bloom filter can be compared and thus, Type-p action fault is detected and localized. Note action fault is Type-p as it is caused in production traffic.

Binary hash chain in \vrd gives \system better accuracy, however, we lose the ability to \emph{automatically} localize the Type-p match faults where the path remains the same and rules matched are different since the \vpd bloom filter remains the same. To summarize, detection will happen always, but localization can happen automatically only in the case of two conditions holding simultaneously:
a) when traffic is production; and b) when there is a change in path since \vpd bloom filter will be different. 
In most cases, fuzz traffic is not permitted in the network. Recall active traffic can be injected from any switch in between a source-destination pair. In such cases, the  $R_f$ will still be detected and can be localized by either manual polling of the switches or hop-by-hop traversal from source to destination. Blackholes 
for critical flows can be detected as \consistencychecker generates an alarm after a chosen time of some seconds if it does not receive any packet pertaining to that flow. 
For localizing silent random packet drops, MAX-COVERAGE~\cite{maxcoverage} algorithm can be implemented on \consistencychecker. 
\section{\system Prototype and Evaluation} 
\label{sec:Imp}
\subsection{Prototype}   
\label{sec:prototype}
\subsubsection{\dpl: \prot Shim Header} \label{sec:DPCI}
We decided to use a 64-bit (8 Byte) shim header on layer-2: \prot.
To ensure sufficient space, we limit the link layer MTU to a maximum of 8,092
Bytes for jumbo frames and 1,518 Bytes for regular frames.  \prot has three
fields, namely:\\
$\bullet$\textbf{\vt:} 16-bit EtherType to signify \prot header.\\
$\bullet$\textbf{\vp:} 32-bit encoding the local inport in a switch.\\
$\bullet$\textbf{\vr:} 16-bit encoding the local rule/s in a switch.

We use a new EtherType value for \vtd to ensure that any layer-2
switch on the path without our OpenFlow modifications will forward the
packet based on its layer-2 information. 
The \prot shim header is inserted within the layer 2 header after the source MAC address, just like a VLAN
header. In presence of VLAN (802.1q), the \prot header is
inserted before the VLAN header. Note \prot header is backward compatible with legacy L2 switches, and transparent to other protocols.\\
\subsubsection{\dpl: New OpenFlow Actions} 
\label{sec:actions}
The new actions ensure that there is no interference in forwarding as no extra rules are added. To ensure efficient
use of the shim header space, we use the bloom filter to encode path-related information in the \vpd field and binary hash chains to encode rule-level information in the \vrd field. A binary hash chain adds a new hash-entry to an existing hash-value by computing a hash
of the existing hash-value and the new hash-entry and then storing it as the
new value. The \vpd field is a Bloom-filter which will contain all intermediate hash results including the first and last value. This ensures that we can test the initial value as well as the final path efficiently.\\ 
$\bullet$\textbf{\textit{set\_\vp}:} Computes hash of the unique identifier (\up) of the switch ID and its inport ID
and adds the result to the Bloom-filter in the \vpd field.\\
$\bullet$\textbf{\textit{set\_\vr}:} Computes hash of the globally unique identifier (\ur) of the flow rule, i.e., switch ID and rule ID
(uniquely identifying a rule within a table), flow table ID with the previous value of the \vrd to form a binary hash chain.\\
$\bullet$\textbf{\textit{push\_verify}:} Inserts a \prot header if needed, initializes the value in \vrd to 1 and the value of \vpd is the hash of \up. It is immediately followed by \textit{set\_\vr} and \textit{set\_\vp}. If there is no \prot header, \textit{push\_verify} is executed at the entry and the exit switch between a source and destination pair.\\
$\bullet$\textbf{\textit{pop\_verify}:} Removes the \prot header from the tagged
packet. 

\textit{push\_verify} should be used, if there is no \prot header for a) all packets entering a source inport or b) all packets leaving the destination outport (in case, if any traffic is injected between a source-destination pair) just before a report is generated to the \consistencychecker. For packets leaving the destination outport, \textit{pop\_verify} should be used only after a
\emph{sampled} report to the \consistencytester has been generated.

To initiate and execute data plane tagging, the actions \textit{set\_\vp} and
\textit{set\_\vr} are prepended to all flow rules in the switches as first actions in the
OpenFlow ``action list''~\cite{onefive}. On the entry and exit switch, action
\textit{push\_verify} is added as the first action. On the exit switch, \textit{pop\_verify} is added as an action once the report is generated. Recall, our actions do not change the forwarding behavior per se as the match
part remains unaffected. However, if one of the actions gets modified
unintentionally or maliciously, it may have a negative impact but gets detected and localized later. Notably, \textit{set\_\vr} encodes the priority of the rule and flow table number in the \vrd field and thus, providing support for rule priorities/cascaded flow tables.
\subsubsection{Bloom Filter \& Hash Function} 
\label{sec:hash}
We use \vpd bloom filter for the localization of detected faults. In an extreme case from the perspective of operational networks like datacenter networks or campus networks, for a packet header and a pair of ingress and egress port, if \cp computes $i$ different paths with $n$ hops in each of the paths, the probability of a collision in bloom filter and hash value simultaneously will be given by: $(0.6185)^{m/n} \times p(i)$

In our case, $(0.6185)^{m/n} = (0.6185)\textsuperscript{32/n}$ using the bloom filter false positive formula~\cite{bloom2} as $m = 32$ (bloom filter size) of the \vpd field and $n$ is the network diameter or the number of switches in a path. $p(i)$ is the probability of collision of the hash function computed using a simple approximation of the birthday attack formula:
\\ $p(i) = i^2/2H = i^2/2^{17}$

$i$ is the number of different paths, $H$ is 2\textsuperscript{16} for 16-bit \vrd field hash. 

For the 16-bit \vrd hash operation, we used Cyclic Redundancy Check (CRC) code~\cite{crc}. For the 32-bit \vpd Bloom filter operations, we use one of the similar approaches as mentioned in~\cite{bloom2}. First, three hashes are computed as: $g_i(x) = h1(x) + i\cdot h2(x)$ for $i = 0, 1, 2$ where $h1(x)$ and $h2(x)$ are the two halves of a 32-bit Jenkins hash~\cite{jenkins} of $x$. Then, we use the first 5 bits of $g_i(x)$ to set the 32-bit Bloom filter for $i = 0, 1, 2$. 

\smartparagraph{\system Components:} We implemented \dpl on top of software switches, in particular, Open vSwitch~\cite{ovs} Version~2.6.90. The customized OvS switches and the fuzz/production traffic generators run in vagrant VMs. Currently, the prototype supports both OpenFlow 1.0 and OpenFlow 1.1. In our prototype, we chose Ryu SDN controller. Python-based \consistencychecker, Java-based \cp and Python-based \fuzz communicate through Apache Thrift~\cite{thrift}. 
\vspace{-.75em}
\subsection{Experiment Setup}
\begin{figure}[!tbp]
	\centering
	\begin{minipage}[b]{0.13\textwidth}
		\centering
		\includegraphics[width=\linewidth]{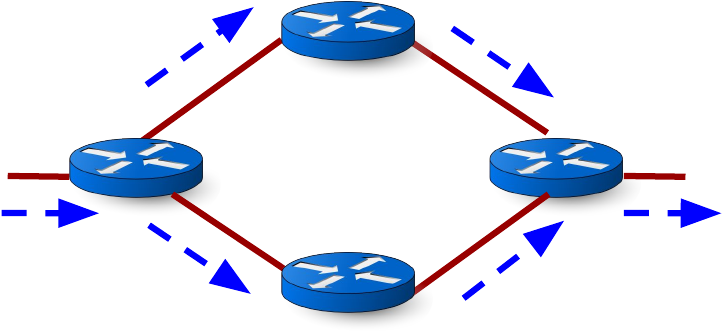}
	\end{minipage}%
	\hfill
	\begin{minipage}[b]{0.15\textwidth}
		\centering
		\includegraphics[width=\linewidth]{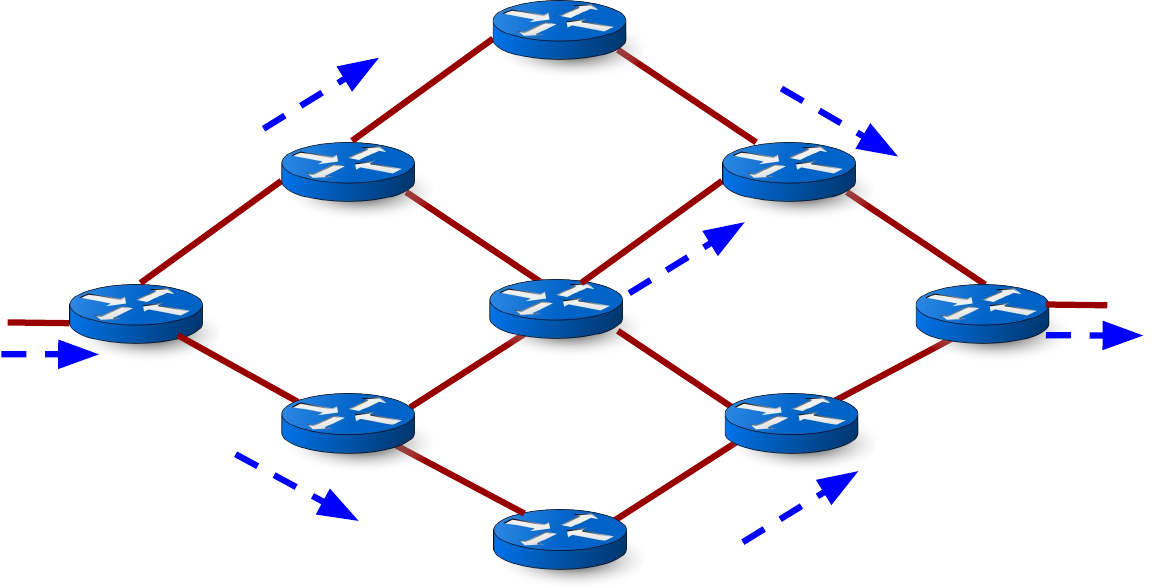}
	\end{minipage}
	\hfill
	\begin{minipage}[b]{0.15\textwidth}
		\centering
		\includegraphics[width=\linewidth]{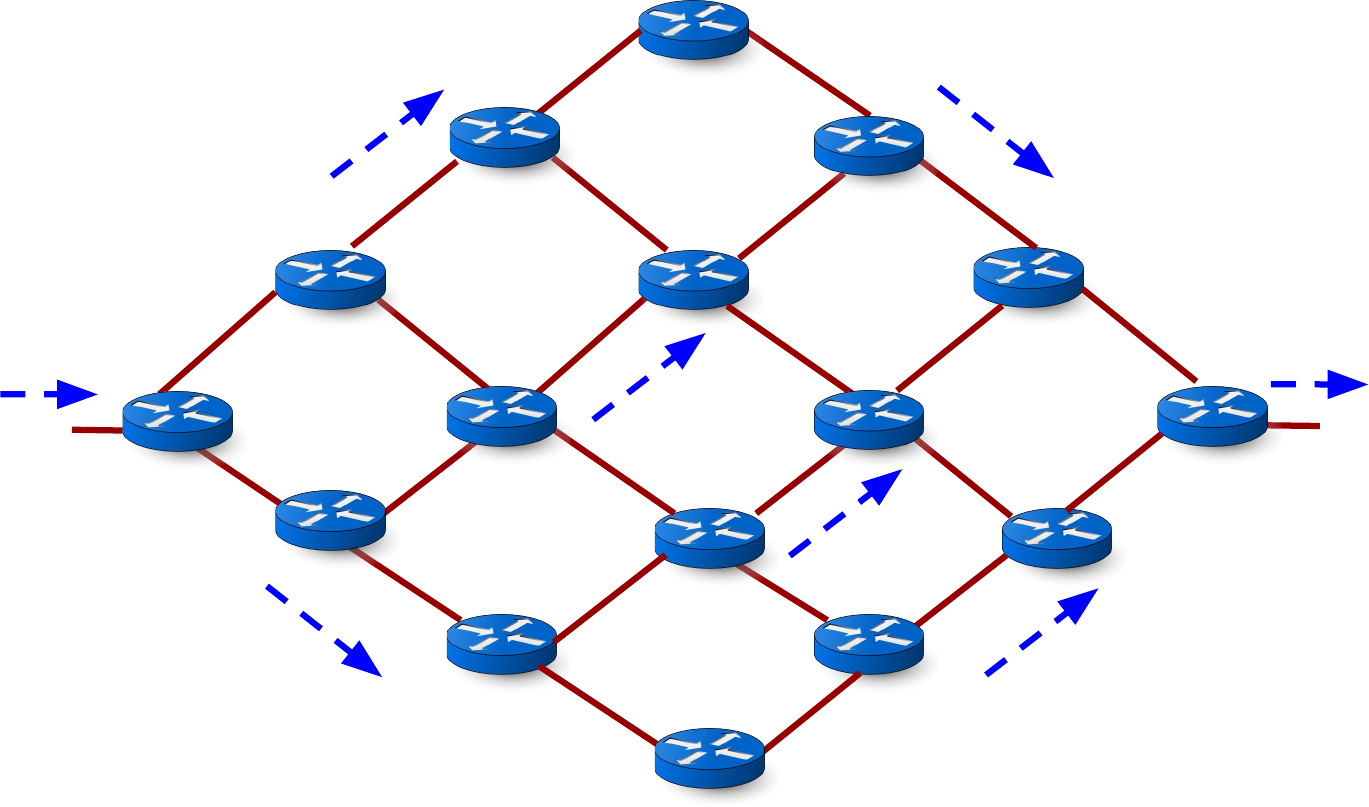}
	\end{minipage}
	\caption{Left to right: Experimental grid topologies with 4, 9 and 16 switches respectively where fuzz and production traffic enter from the leftmost switch and leave at the rightmost switch in each topology.}
	\label{fig:test}
\end{figure}
\begin{figure}[tbp]
	\centering
	\includegraphics[width=0.8\columnwidth]{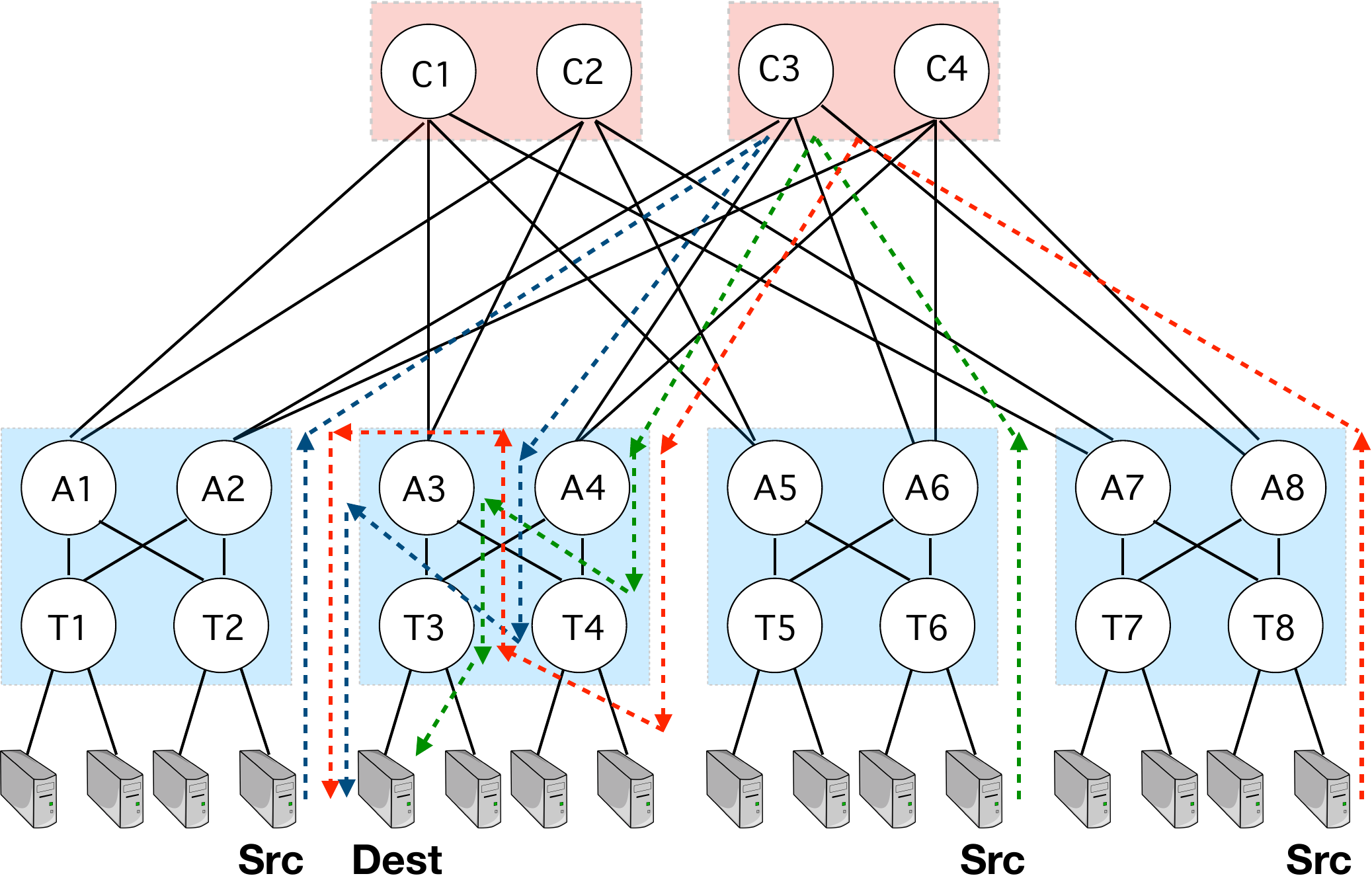}
	\caption{4-ary fat-tree (20 switches) experimetal topology with expected paths in dotted blue line from three different sources (Src) to Destination (Dst). T$<x>$: ToR (Top-of-Rack), A$<x>$: Aggregate and C$<x>$: Core switches respectively. Paths for each source-destination pair: T2 A2 C3 A4 T4 A3 T3 (Blue), T6 A6 C3 A4 T4 A3 T3 (Green), T8 A8 C3 A4 T4 A3 T3 (Red).}
	\label{fig:fat}
	\vspace{-.5em}
\end{figure}

We evaluate \system on 4 topologies: a) 3 grid topologies (Figure~\ref{fig:test}) 
of 4, 9 and 16 switches respectively with varying complexities to ensure diversity of paths, and b) 1 datacenter fat-tree (4-ary) topology of 20 switches with multipaths (Figure~\ref{fig:fat}). 
Experiments were conducted on an 8 core 2.4GHz Intel-Xeon CPU machine and 64GB of RAM. For scalability purposes, we modified and translated the Stanford backbone configuration files~\cite{1_peymank} to equivalent OpenFlow rules as per our topologies, and installed them at the switches to allow multi-path destination-based routing. We used our custom script to generate configuration files for the four experimental topologies. The configuration files ensured the diversity of paths for the same packet header. Columns 1-4 in Table~\ref{tab:sta} illustrate the parameters of the four experimental topologies.
\begin{table}[t]
	\centering
	{\small
		\scalebox{0.74}{	
			\begin{tabular}{|m{2.4cm}|m{0.7cm}|m{0.7cm}|m{1cm}|m{2cm}|m{2cm}|}
				\hline
				Topology& \#Rules& \#Paths& Path Length & Reachability graph computation time & \fuzz Execution Time\\
				\hline
				4 switches (grid)& \textasciitilde5k& \textasciitilde24k & 2 & 0.64 seconds& \textasciitilde 1 ms\\
				\hline
				9 switches (grid)& \textasciitilde27k & \textasciitilde50k& 4& 0.91 seconds& \textasciitilde 1.2 ms\\
				\hline
				16 switches (grid)& \textasciitilde60k& \textasciitilde75k & 6& 1.13 seconds& \textasciitilde 3.2 ms\\
				\hline
				4-ary fat-tree (20 switches)& \textasciitilde100k& \textasciitilde75k & 6& 1.15 seconds& \textasciitilde 7.5 ms\\
				\hline
		\end{tabular}}
		\caption{Columns 1-4 depict the parameters of four experimental topologies.\\
			Column 5 depicts the reachability graph computation time by the \cp for the experimental topologies proactively by the \cp. Represents an average over 10 runs.\\
			Column 6 depicts the \fuzz execution time to compute the packet header space for generating the fuzz traffic for the corresponding experimental topologies. Represents an average over 10 runs.}
		\label{tab:sta}		
		\vspace{-2em}
	}
\end{table}
\begin{figure*}[t]
	\captionsetup[subfigure]{justification=justified,singlelinecheck=false}
	\centering
	\begin{minipage}[b]{0.5\linewidth}
		\begin{subfigure}[b]{0.3\linewidth}
			\centering
			\includegraphics[width=\linewidth]{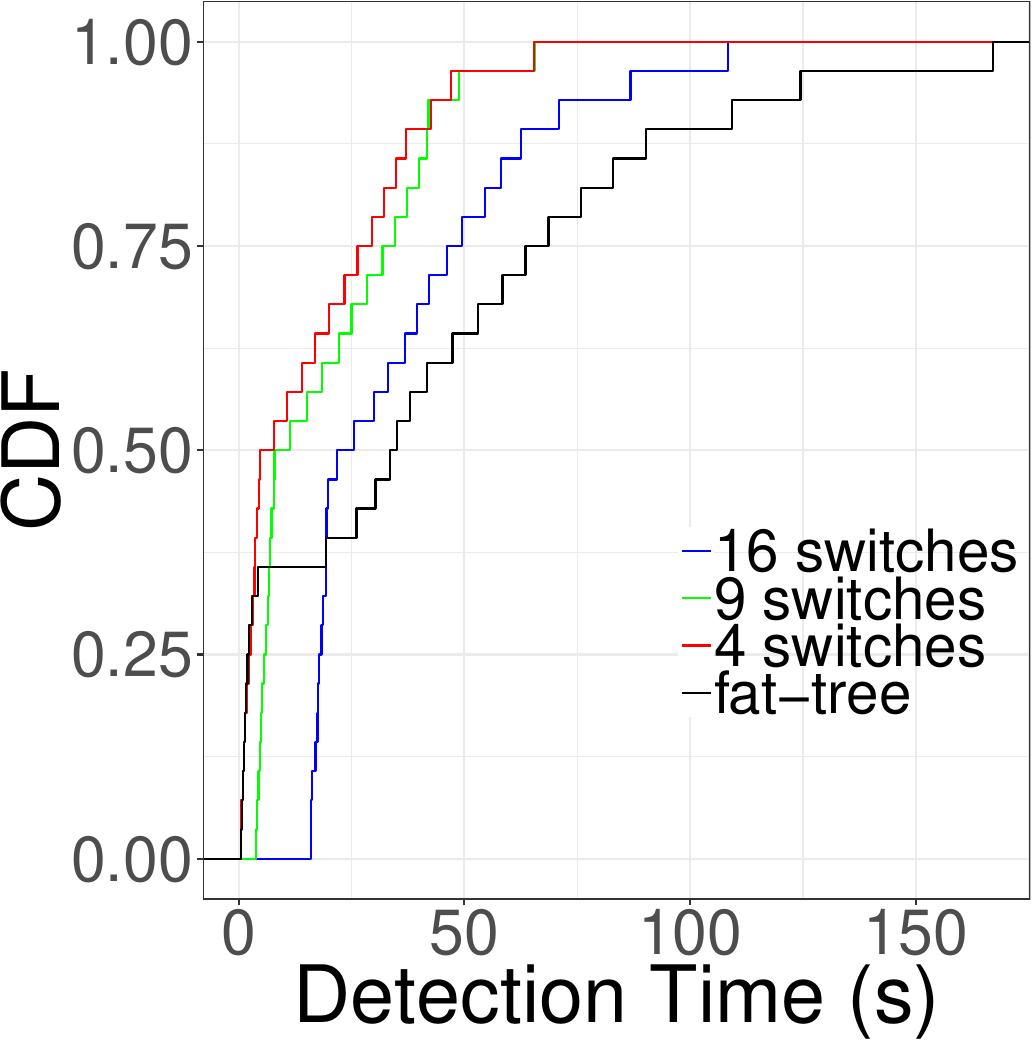}
		\end{subfigure}
		\begin{subfigure}[b]{0.3\linewidth}
			\centering
			\includegraphics[width=\linewidth]{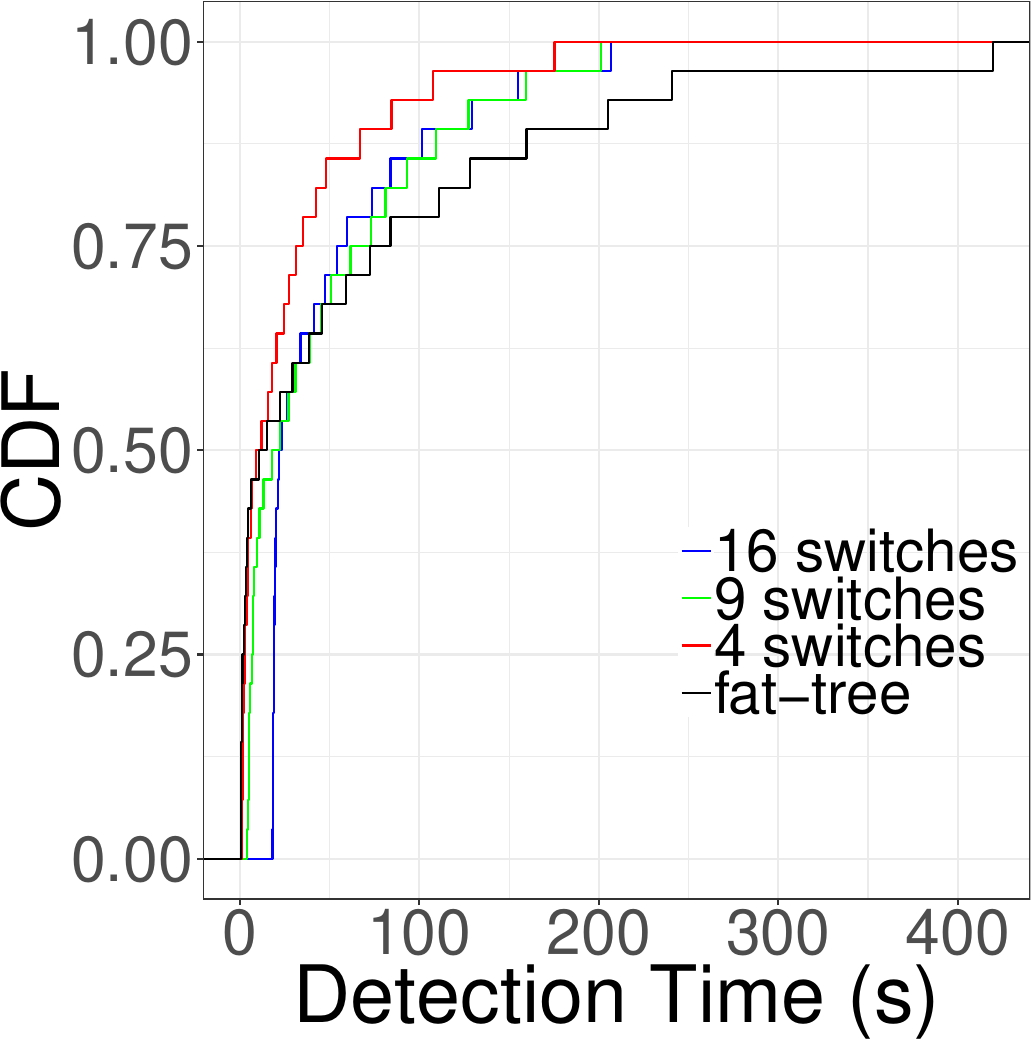}
		\end{subfigure}
		\subcaption{Results of detection time for all topologies}
		\label{fig:plots}
	\end{minipage}%
	\hspace*{-3cm}
	\begin{minipage}[b]{0.5\linewidth}
		\begin{subfigure}[b]{0.3\linewidth}
			\centering
			\includegraphics[width=\linewidth]{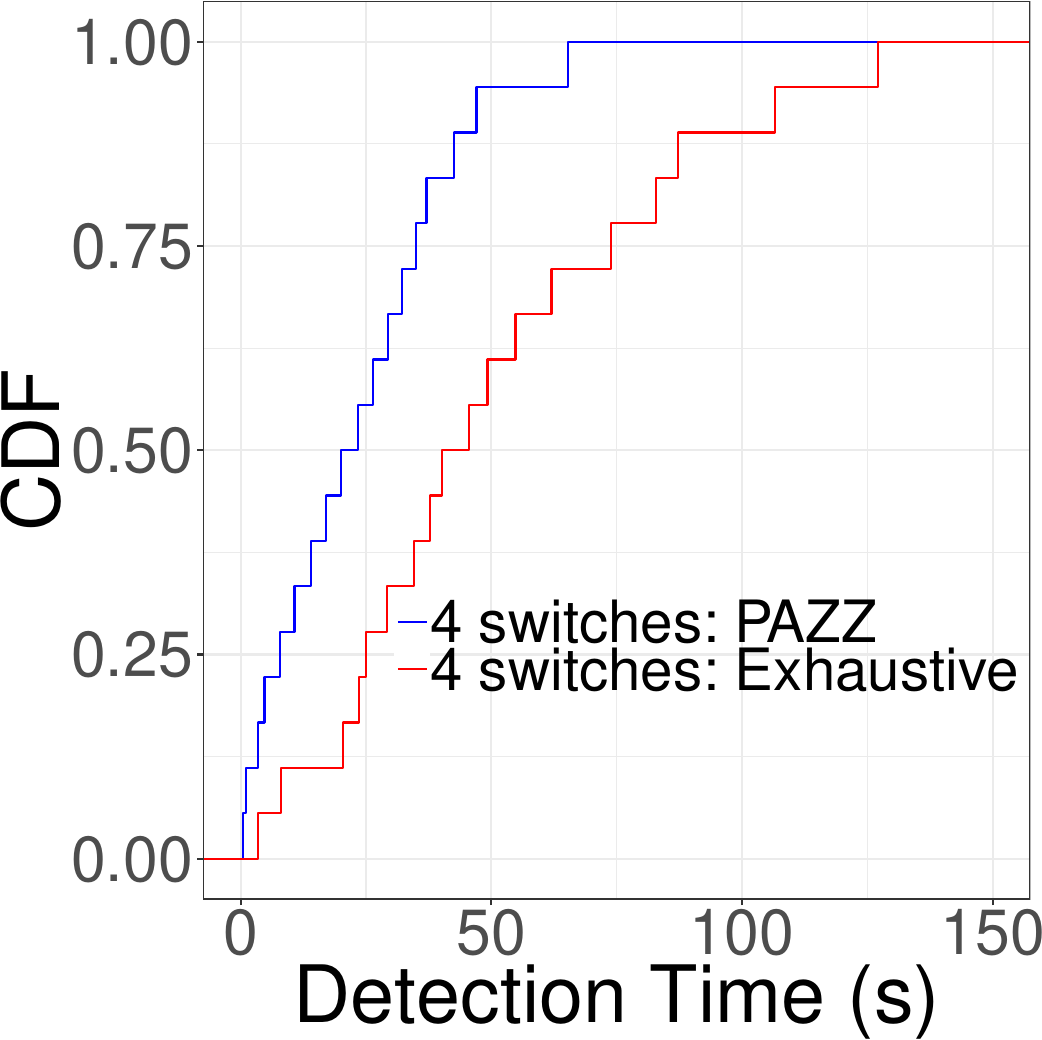}
		\end{subfigure}
		\begin{subfigure}[b]{0.3\linewidth}
			\centering
			\includegraphics[width=\linewidth]{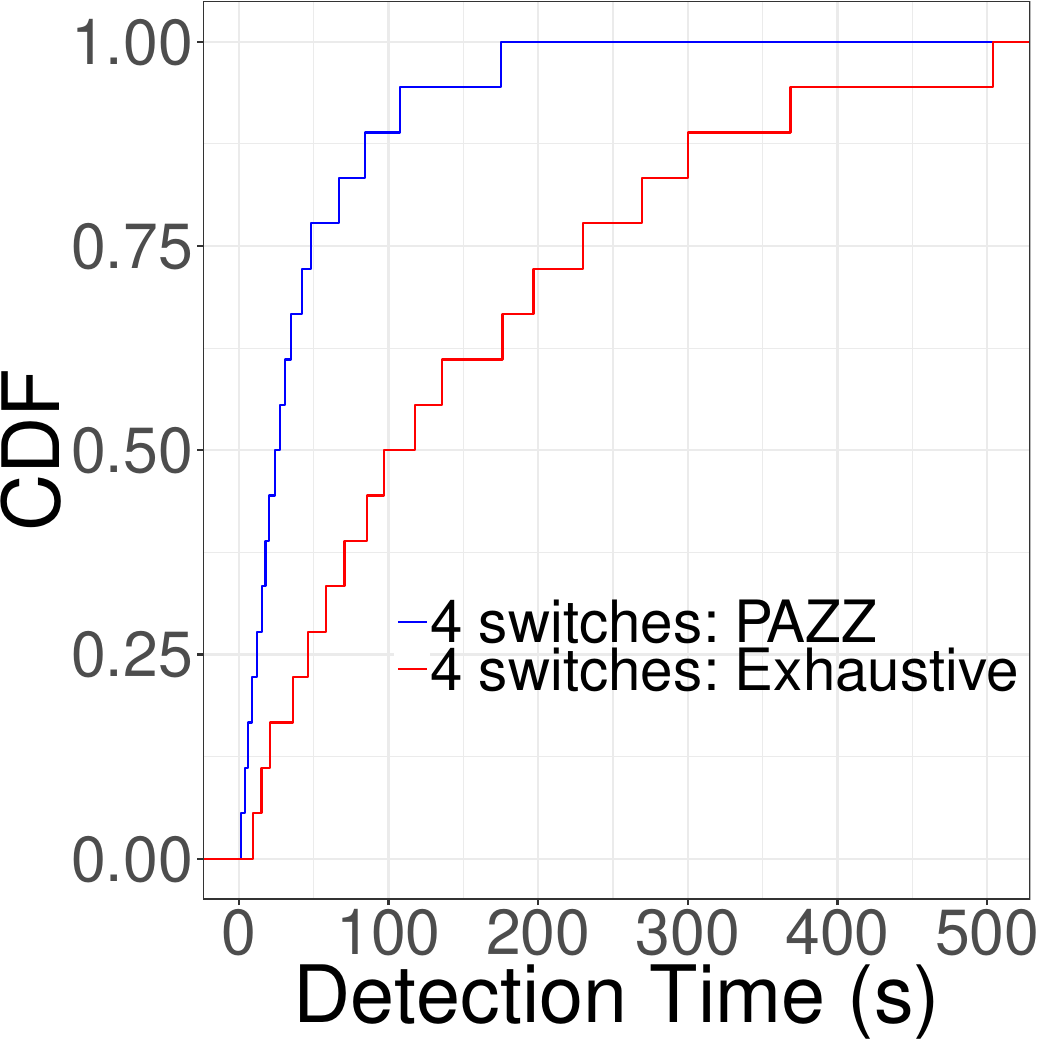}
		\end{subfigure}
		\subcaption{Results of 4-switch grid topology}
		\label{fig:4comp}
	\end{minipage}%
	\hspace*{-3cm}
	\begin{minipage}[b]{0.5\linewidth}
		\begin{subfigure}[b]{0.3\linewidth}
			\centering
			\includegraphics[width=\linewidth]{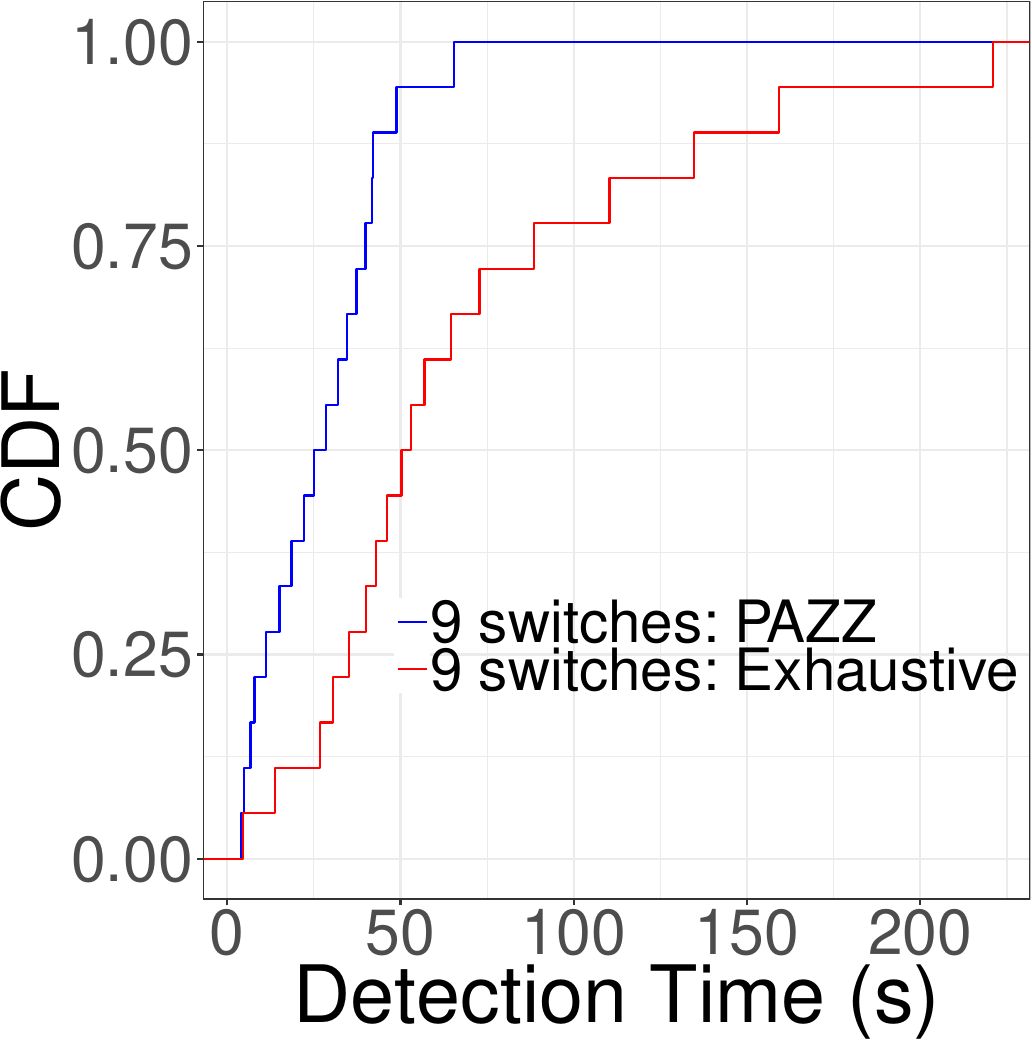}
		\end{subfigure}
		\begin{subfigure}[b]{0.3\linewidth}
			\centering
			\includegraphics[width=\linewidth]{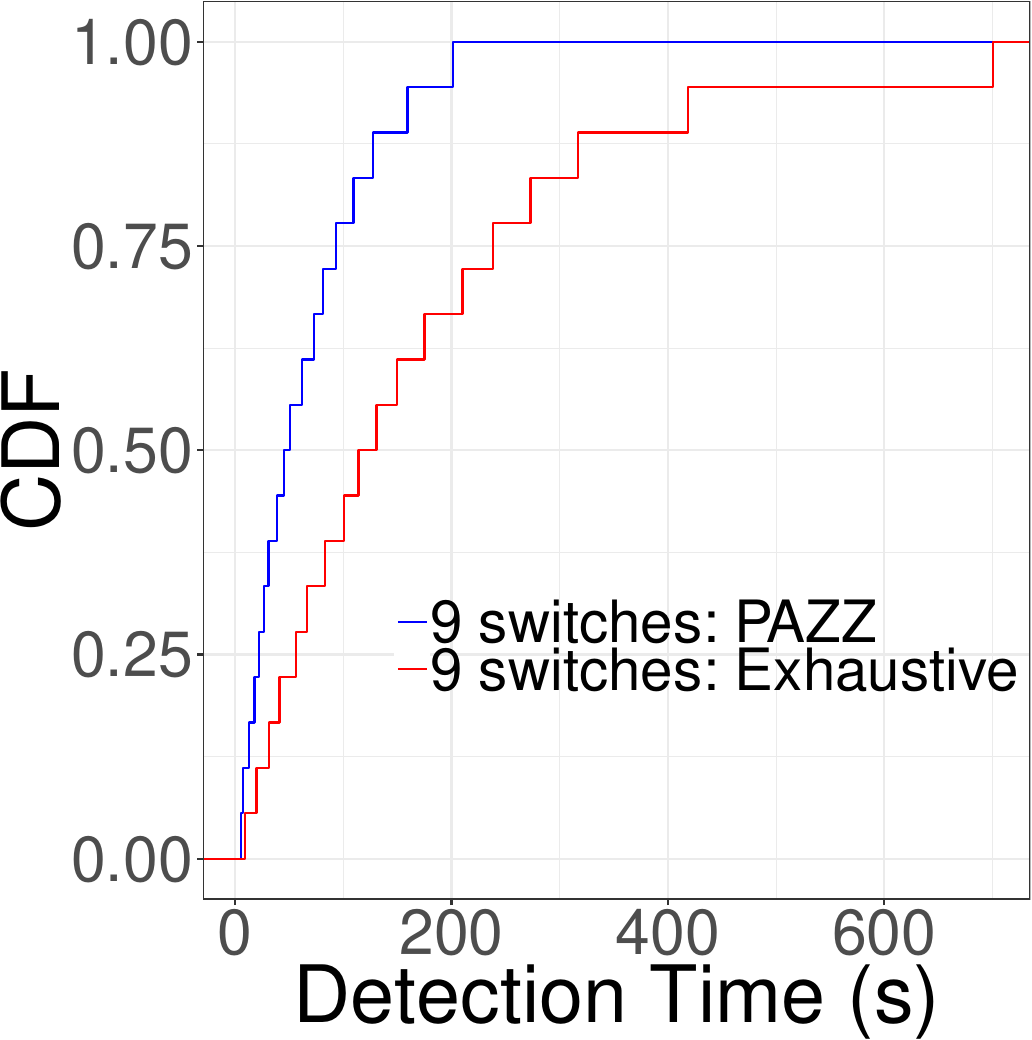}
		\end{subfigure}
		\subcaption{Results of 9-switch grid topology}
		\label{fig:9comp}
	\end{minipage}%
	\hfill
	\begin{minipage}[b]{0.5\linewidth}
		\begin{subfigure}[b]{0.3\linewidth}
			\centering
			\includegraphics[width=\linewidth]{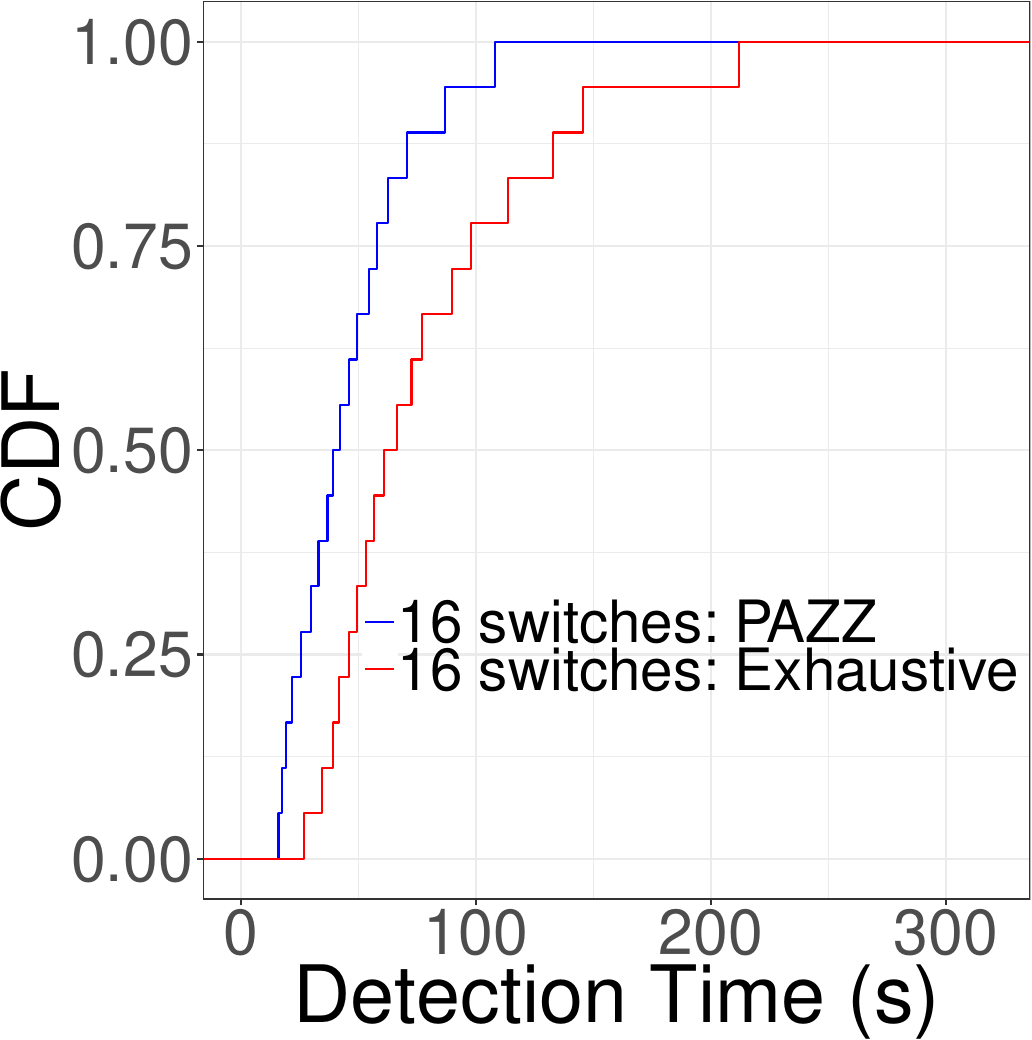}
		\end{subfigure}
		\begin{subfigure}[b]{0.3\linewidth}
			\centering
			\includegraphics[width=\linewidth]{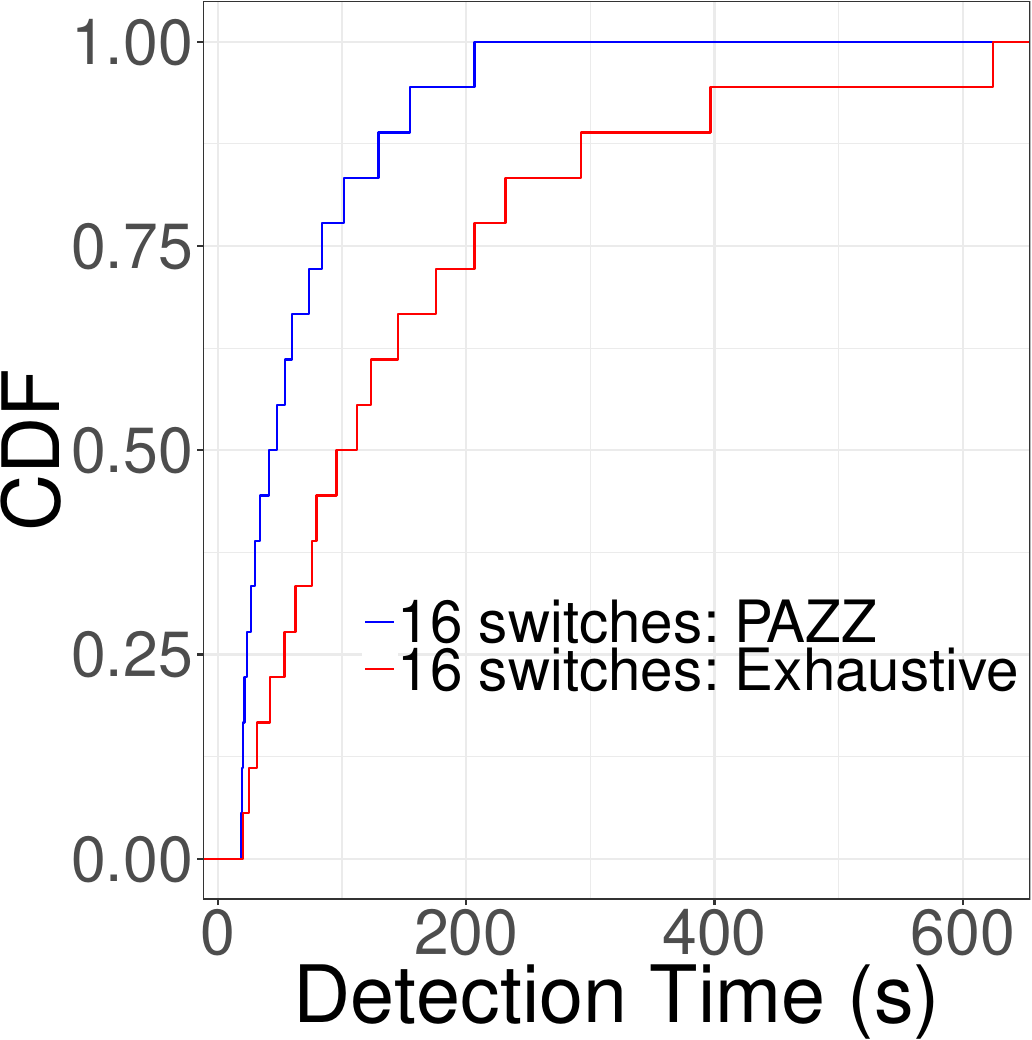}
		\end{subfigure}
		\subcaption{Results of 16-switch grid topology}
		\label{fig:16comp}
		\vspace{-.25em}
	\end{minipage}%
	\hspace*{-3cm}
	\begin{minipage}[b]{0.5\linewidth}
		\begin{subfigure}[b]{0.3\linewidth}
			\centering
			\includegraphics[width=\linewidth]{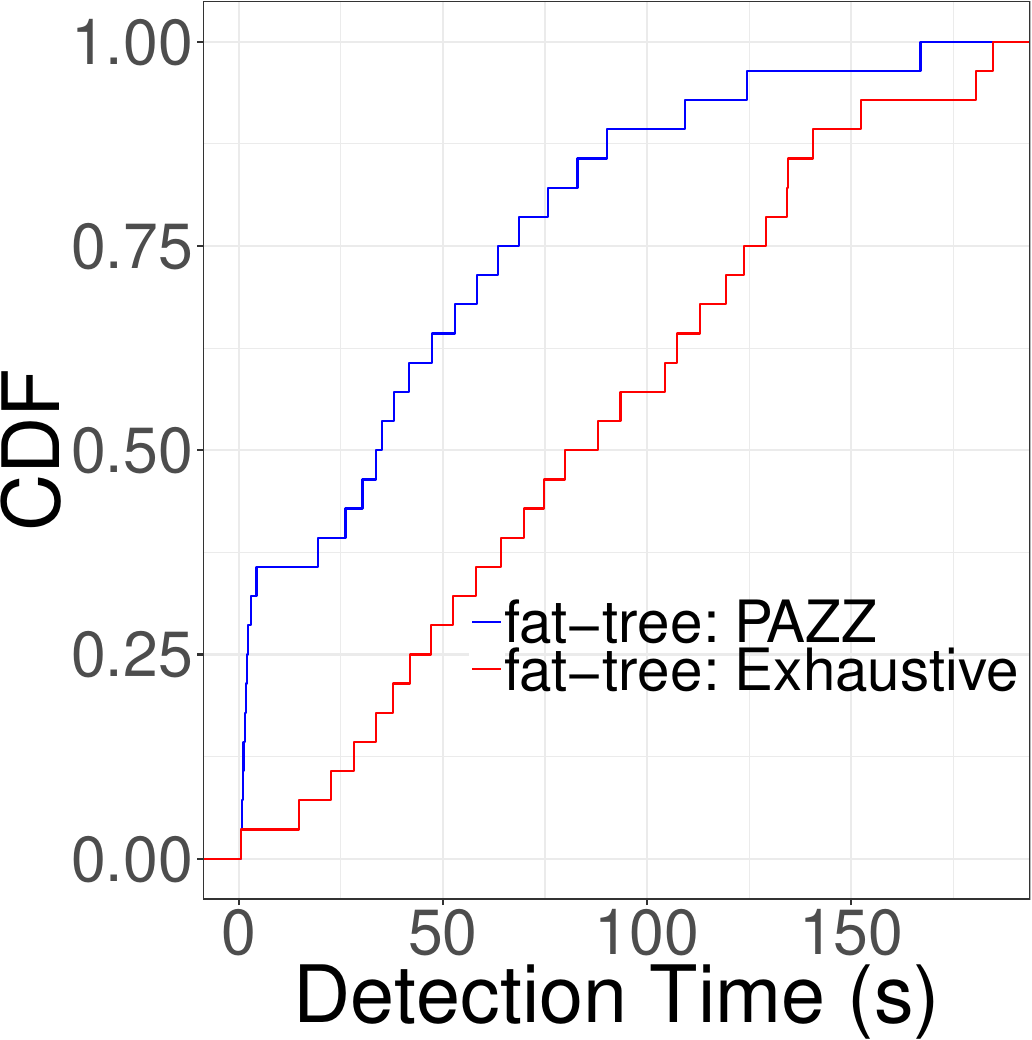}
		\end{subfigure}
		\begin{subfigure}[b]{0.3\linewidth}
			\centering
			\includegraphics[width=\linewidth]{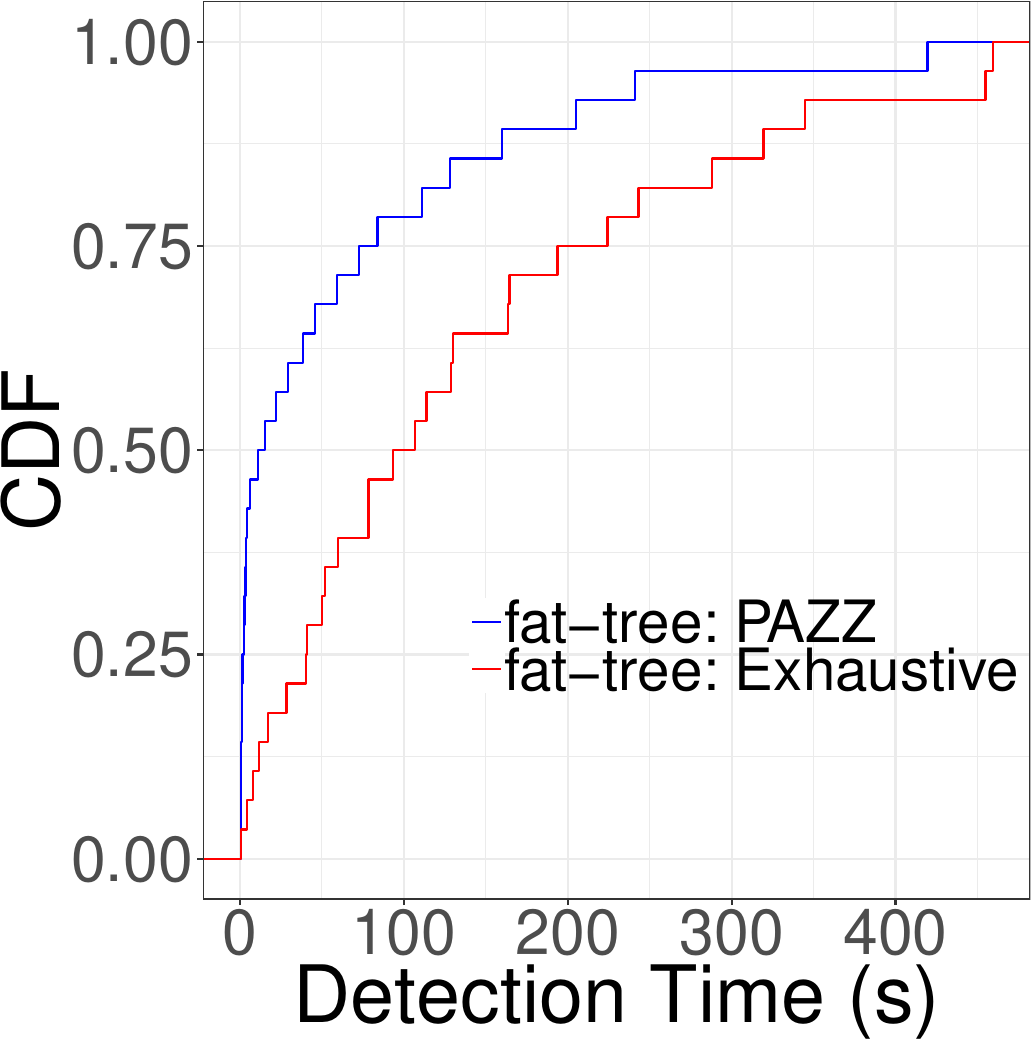}
		\end{subfigure}
		\subcaption{Results of 4-ary fat-tree (20-switch) topology}
		\label{fig:fatcomp}
		\vspace{-.25em}
	\end{minipage}%
	\hspace*{-3cm}
	\begin{minipage}[b]{0.5\linewidth}
		\begin{subfigure}[b]{0.3\linewidth}
			\centering
			\includegraphics[width=\linewidth]{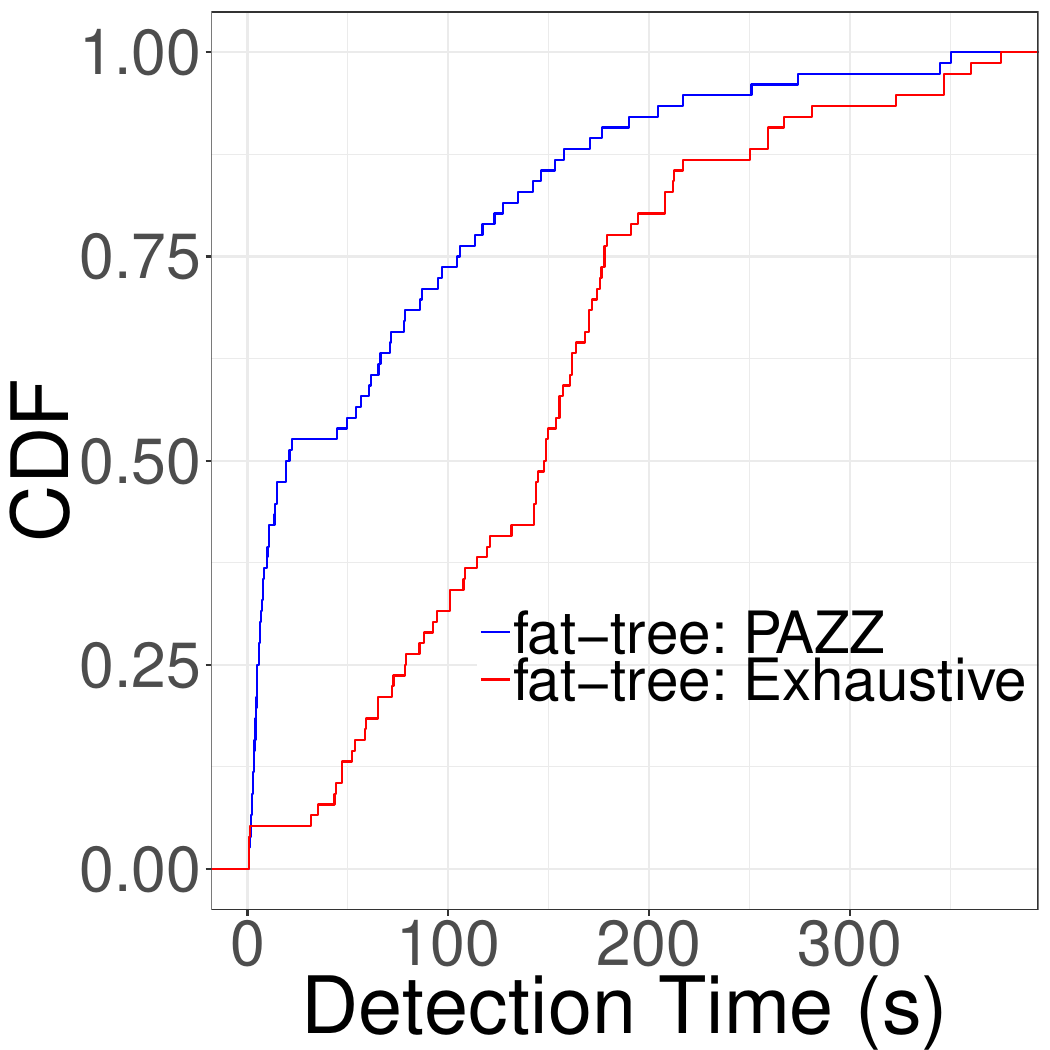}
		\end{subfigure}
		\begin{subfigure}[b]{0.3\linewidth}
			\centering
			\includegraphics[width=\linewidth]{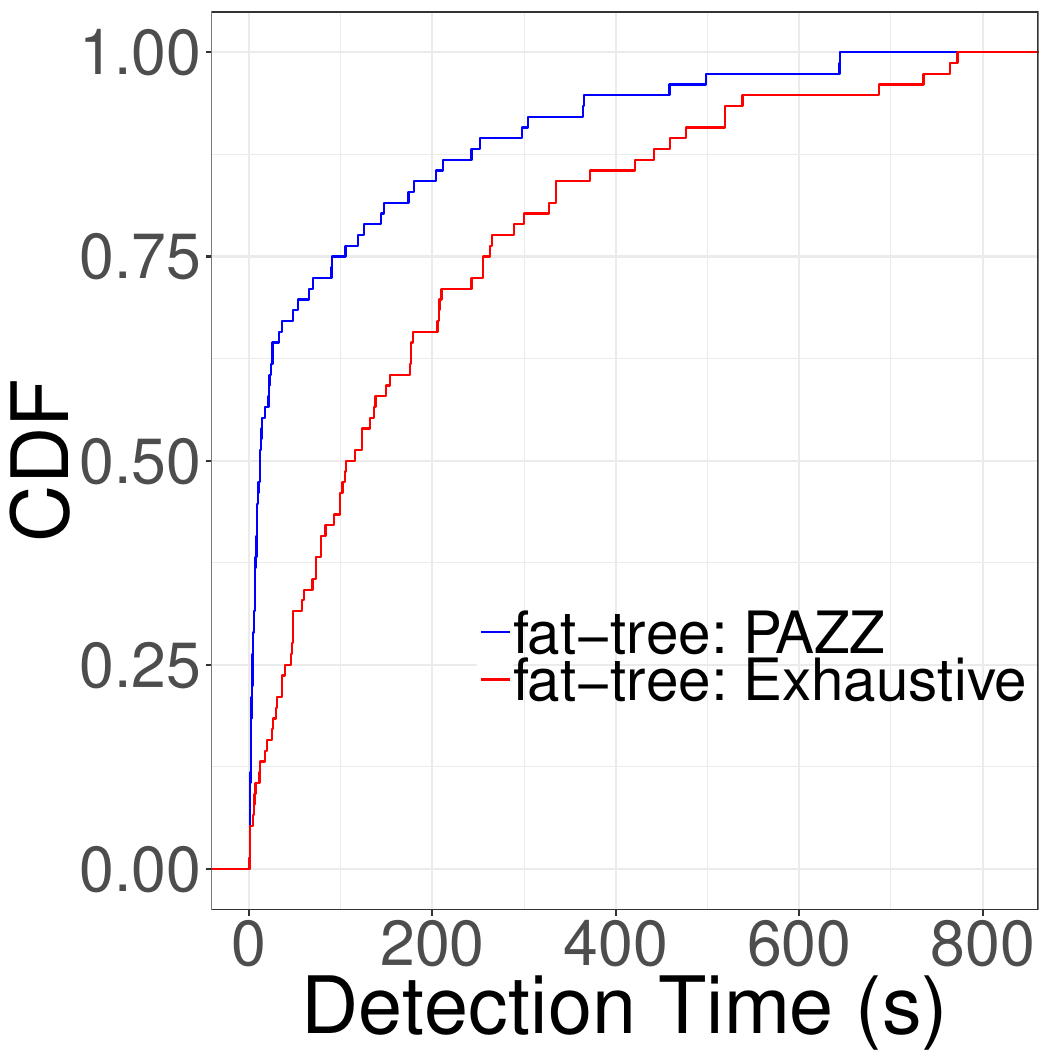}
		\end{subfigure}
		\subcaption{Results of 3 source-destination pairs in fat-tree}
		\label{fig:3fatcomp}
		\vspace{-.25em}
	\end{minipage}%
	\caption{a) For a source-destination pair, CDF of Type-p and Type-a fault detection time by \system in all 4 experimental topologies for sampling rate of 1/100 (left), 
		and 1/1000 (right) respectively. 
		Red, green, blue and black lines belong to 4-switch, 9-switch, 16-switch and 4-ary fat-tree topology (20-switch) respectively. \newline
		b), c), d), e) For a source-destination pair, comparison of fault detection time by \system (blue) and exhaustive packet generation approach (red) in all 4 experimental topologies (4-switch, 9-switch, 16-switch and 4-ary fat-tree respectively). In each figure, left to right illustrates sampling rate of 1/100 (left), 
		and 1/1000 (right) respectively. \newline
		f) For 3 source-destination pairs, comparison of fault detection time (in seconds) by \system (blue) and exhaustive packet generation approach (red) in 4-ary fat-tree topology. Left to right illustrates sampling rate of 1/100 (left), 
		and 1/1000 (right) respectively.} \label{fig:total} 
	\vspace{-.75em}
\end{figure*}
\begin{figure*}[t]
	\centering
	\begin{subfigure}[b]{0.5\linewidth}
		\centering
		\includegraphics[width=\linewidth]{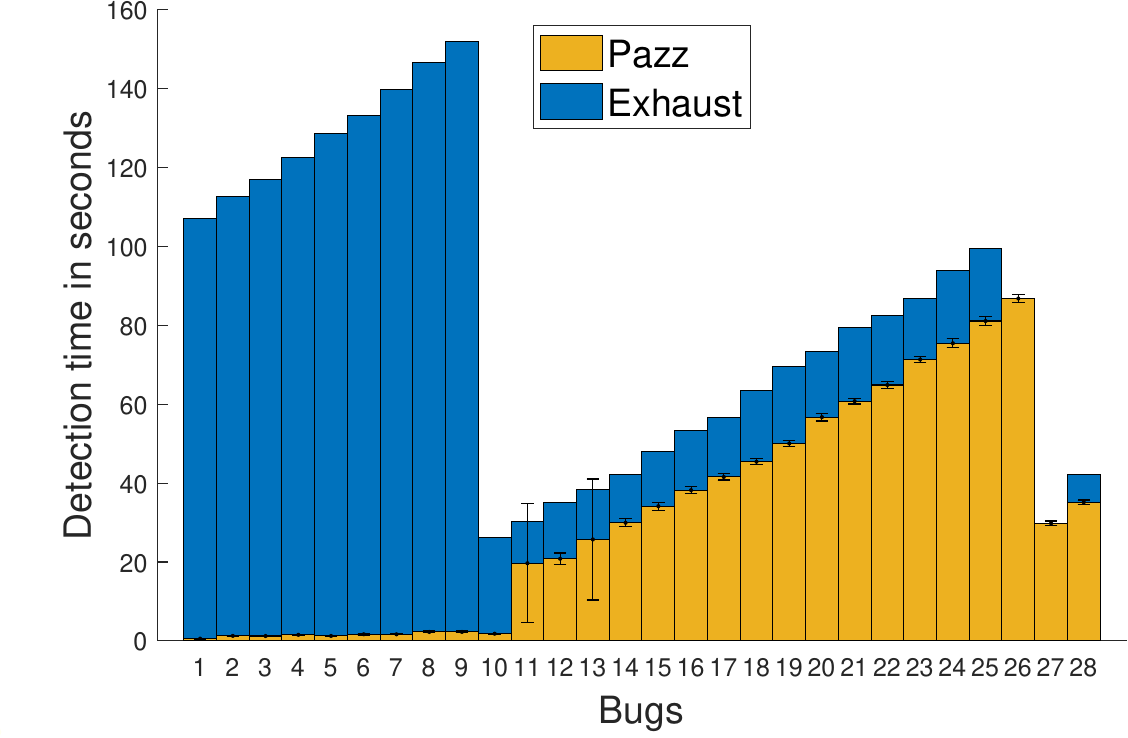}
		\subcaption{Results of 4-ary fat-tree (20-switch) topology with 1/100 sampling rate}\label{spectrum1}
	\end{subfigure}%
	\begin{subfigure}[b]{0.5\linewidth}
		\centering
		\includegraphics[width=\linewidth]{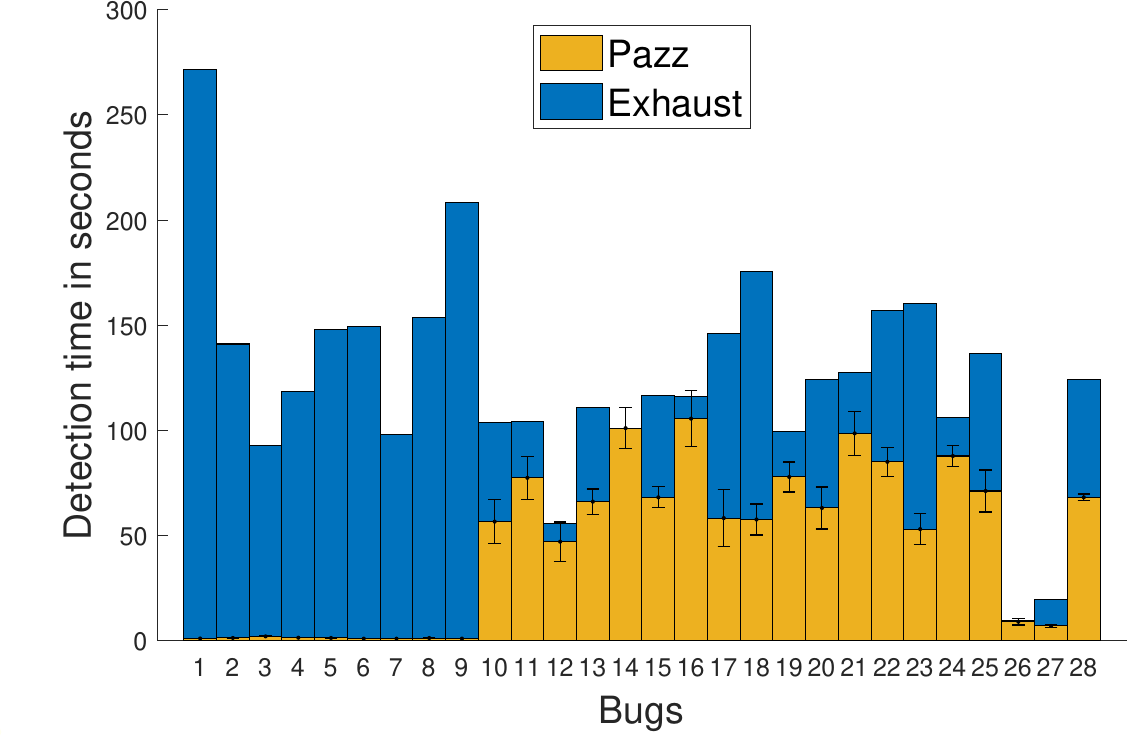}
		\subcaption{Results of 4-ary fat-tree (20-switch) with 1/1000 sampling rate}\label{spectrum2}
	\end{subfigure}%
	\caption{a) Illustrates the median fault detection time in \system with quartiles vs baseline for a single source-destination pair in 4-ary fat-tree topology with sampling rate of 1/100. (Figure~\ref{fig:fatcomp} (left) represents CDF of the mean fault detection time.)\\
		b) Illustrates the median fault detection time in \system with quartiles vs baseline for a single source-destination pair in 4-ary fat-tree topology with sampling rate of 1/1000. (Figure~\ref{fig:fatcomp}~(right) represents CDF of the mean fault detection time.)}
	\label{fig:spectrum}
	\vspace{-1em}
\end{figure*}
We randomly injected faults on randomly chosen OvS switches in the data plane where each fault belonged to different packet header space (in 32-bit destination IPv4 address space) either in the production or fuzz traffic header space. In particular, we injected Type-p (match/action faults) and Type-a faults. \texttt{ovs-ofctl} utility was used to insert the faults in the form of high-priority flow rules on random switches. Therefore, we simulated a scenario where the control plane was unaware of these faults in the data plane. We made a pcap file of the production traffic generated from our Python-based script that crafts the packets. In addition, we made a pcap of the fuzz traffic generated from the \fuzz. The production and fuzz traffic pcap files were collected using Wireshark and replayed  in parallel and continuously at the desired rate using Tcpreplay~\cite{turner2005tcpreplay} with infinite loops to test the network continuously. To execute \system periodically instead of continuously, there is an API support to implement timers with desired timeouts as per the deployment scenario. For sampling, we used sFlow~\cite{sflow} with a polling interval of 1 second and a sampling rate of 1/100, 1/500 and 1/1000. Note, the polling interval and sampling rate can be modified as per the deployment scenario. The sampling was done on the egress port of the exit switch in the data plane so the sampled actual report reaches the \consistencychecker and thus, avoids overwhelming it. Note each experiment was conducted for ten times for a randomly chosen source-destination pair. 

\smartparagraph{Workloads:} For 1 Gbps links between the switches in the 4 experimental topologies: 3 grid and 1 fat-tree (4-ary), the production traffic was generated at 10\textsuperscript{6} pps (packets per second). In parallel, fuzz traffic was generated at 1000 pps, i.e., $0.1$\% of the production traffic. Note, the fuzz traffic rate can be modified as per the use-case requirements.
\subsection{Evaluation Strategy}
For a source-destination pair, our experiments are parameterized by: (a) size of network (4-20 switches), (b) path length (2-6), (c) configs (flow rules from 5k-100k), (d) number of paths (24k-75k), (e) number around (1-30) and kind of faults (Type-p, Type-a), (f) sampling rate (1/100, 1/500, 
1/1000) with polling interval (1 sec), and (g) workloads, i.e., throughput (10\textsuperscript{6} pps for production and 1000 pps for fuzz traffic). Note, due to space constraints, we removed the results of 1/500 sampling rate. Our primary metrics of interest are fault detection with localization time, and comparison of fault detection/localization time in \system against the baseline of exhaustive traffic generation approach and Header Space Analysis (HSA)~\cite{Kazemian2012}. 
In particular, we ask these questions:\\
\emph{\textbf{Q1.} How does} \system \emph{perform under different topologies and configs of varying scale and complexity?} (\cref{sec:I1})\\
\emph{\textbf{Q2.} How does} \system \emph{compare to the strawman case of exhaustive random packet generation for a single and multiple source-destination pairs?} (\cref{sec:I2})\\
\emph{\textbf{Q3.} How much time does} \system \emph{take to compute reachability graph as compared to Header Space Analysis~\cite{Kazemian2012}?} (\cref{sec:I4})\\
\emph{\textbf{Q4.} How much time does} \system \emph{take to generate active traffic for a source-destination pair and how much overhead does} \system \emph{incur on the links?} (\cref{sec:I5})\\
\emph{\textbf{Q5.} How much packet processing overhead does} \system \emph{incur on varying packet sizes?} (\cref{sec:I6})\\
\vspace{-1.5em}
\subsection{\system Performance}
\label{sec:I1}

Figure~\ref{fig:plots} illustrates the cumulative distribution function (CDF) of the Type-p and Type-a faults detected in the four different experimental topologies with the parameters mentioned in Table~\ref{tab:sta}. As expected, in a grid 16-switch topology with 60k rules and 75k paths, \system takes only 25 seconds to detect 50\% of the faults and 105 seconds to detect all of the faults in case of sampling rate 1/100 and polling interval of 1 second (left in Figure~\ref{fig:plots}). For the same sampling rate of 1/100, in the case of 4-ary fat-tree topology with 20 switches containing 100k rules and 75k paths, \system detects 50\% of the faults in 40 seconds and all faults in 160 seconds. Since the production traffic was replayed at 10\textsuperscript{6} pps in parallel with the fuzz traffic replayed at 1000 pps, the Type-p faults in the production traffic header space (35\% of total faults) were detected faster in a maximum time of 24 seconds for all four topologies as compared to the Type-a faults (65\% of total faults) in the fuzz traffic header space which were detected in a maximum time of 420 seconds.
As the experiment was conducted ten times, the time taken is the mean of the ten values to detect a fault pertaining to a packet header space. We omitted confidence intervals as they are small after 10 runs. In all cases, the detection time difference was marginal. \\
\smartparagraph{Localization Time:} As per Algorithm~\ref{alg:loc}, the production traffic-specific faults after detection were automatically localized within a span of 50 $\mu$secs for all four experimental topologies. The localization of faults pertaining to fuzz traffic was manual as there was no expected report from the \cp. Hereby, the localization was done for two cases: a) when the fuzz traffic entered at the ingress port of the entry switch and b) when the fuzz traffic entered in between a pair of ingress and egress ports. For the first case, the localization of each fault happens in a second after the fault was detected by the \consistencychecker as the first switch possessed a flow rule to allow such traffic in the network. For the second case, i.e., where fuzz traffic was injected from between the pair of ingress and egress ports took approx. 2-3 minutes after detection for manual localization as the path was constructed after hop-by-hop inspection of the switch rules.

To compute the CPU usage in \system, we measured the CPU time ($cpu\_time = cpu\_cycles * cpu\_clock\_rate$) required for computing expected paths, test and detect inconsistency per sFlow sample analyzed by the \ct (brain of \system) for the 4-ary fat-tree scenario. Our single threaded implementation, required $4.271$ ms (mean), and $4.397$ ms (median) with a minimum of $0.259$ ms, and a maximum of $7.310$ ms for each sample on a single core of the Intel Xeon E$5$-$2609$ $2.40$GHz CPU. For the memory usage, topology and configuration are the main factors. The fat-tree scenario with $100$k rules requires the maximum resident set size (RSS) of $1.29$MB. Memory usage of \cp is mentioned in~\cref{sec:I4}.
\vspace{-.75em}
\subsection{Comparison to Exhaustive Packet Generation}
\label{sec:I2}

We compare the fault detection time of \system which uses \fuzz against exhaustive packet generation approach. For a fair comparison, the exhaustive packet generation approach generates the same number of flows \emph{randomly} and at the same rate like \system. Figures~\ref{fig:4comp}, ~\ref{fig:9comp}, ~\ref{fig:16comp} and~\ref{fig:fatcomp} illustrate the fault detection time CDF in 4-switch, 9-switch, 16-switch and 4-ary fat-tree (20-switch) experimental topologies respectively. Two figures for each experimental topology illustrates the results for two different sampling rates of 1/100, 
and 1/1000 (left and right) respectively. The blue line indicates \system which uses \fuzz and the red line indicates exhaustive packet generation approach. As expected, we observe that \system performs better than exhaustive packet generation approach. \system provides an average speedup of 2-3 times. We observe in all cases, 50\% of the faults are detected in a maximum time of \textasciitilde50 seconds or less than a minute by \system. Note we excluded the \fuzz execution time (\cref{sec:I5}) in the plots. It is worth mentioning that \system will perform much better if we compare against a fully exhaustive packet generation approach which generates 2\textsuperscript{32} flows in all possible destination IPv4 header space. Hereby, the detected faults are Type-a as they require active probes in the \emph{uncovered} packet header space. Since \system relies on production traffic to detect the Type-p faults hence, we get rid of the exhaustive generation of all possible packet header space. Similar results were observed for localization of detected faults. 

\smartparagraph{Multiple Source-Destination Pairs \& Header Coverage:}
We observed the similar results for different source-destination pairs when we placed \fuzz and production traffic generator at different sources in parallel for a different header space coverage. Figure~\ref{fig:3fatcomp} shows the fault detection time comparison of \system against exhaustive packet generation approach for 3 different source-destination pairs in the 4-ary fat tree topology (Figure~\ref{fig:fat}). Note3, \system can be extended to Layer-2 and Layer-4 headers.

Figures~\ref{spectrum1},~\ref{spectrum2} illustrate the median fault detection time with quartiles in \system against the baseline for 4-ary fat-tree topology (20 switches) with 100k rules for a single source-destination pair for sampling rates: 1/100 and 1/1000 respectively. Note, due to space constraints, we removed the quartile plots for the three grid topologies. To summarize, the results of the four experimental topologies are as expected and similar without any observable deviation.

\vspace{-.75em}
\subsection{Reachability Graph Computation vs Header Space Analysis (HSA)~\cite{Kazemian2012}}
\label{sec:I4}
Table~\ref{tab:sta} (Column 5) shows the reachability graph computation in all experimental topologies by the \cp for an egress port. To observe the effect of evolving configs, we added additional rules to various switches randomly. We observe that \cp computes the reachability graph in all topologies in $<1$s.

We compare \cp against the state-of-the-art Header Space Analysis (HSA)~\cite{Kazemian2012} that can be used to generate reachability graphs to generate test packets and compute covered header space. HSA is also used by NetPlumber~\cite{Kazemian2013} and ATPG~\cite{Zeng2014}. We observe that HSA library has scalability issues due to poor support for set operations on wildcards unlike BDDs~\cite{Bryant:1986:GAB:6432.6433}. Indeed, we reconfirm this by running a simple reachability experiment between a source-destination pair in the fat-tree topology of 20 switches with 100k rules in total. In this experiment, we compare the resource usage of the Python implementation of HSA library with the \cp of \system which uses BDDs. HSA requires in excess of $206$GB of memory to compute only $10$ possible paths between two ports, while the \cp requires only $1.5$GB of memory to compute all 75k paths.
\vspace{-.75em}
\subsection{\fuzz Execution Time \& Overhead}
\label{sec:I5}
\smartparagraph{Execution Time:} Table~\ref{tab:sta} (Column 6) illustrates the time taken by \fuzz to compute the packet header space for fuzz traffic in the four experimental topologies after it receives the covered packet header space (corpus) from the \cp. Since we considered destination-based routing hence, the packet header space computation was limited to 32-bit destination IPv4 address space in the presence of wildcarded rules. 
When some of the rules were added to the data plane, the \cp recomputed the corpus, the new corpus was sent to the \fuzz which recomputed the new fuzz traffic within a maximum time of 7.5 milliseconds.\\
\smartparagraph{Overhead:} The fuzz traffic contains 54-byte test packets at the rate of 1000 pps on a 1 Gbps link that is 0.04\% of the link bandwidth and therefore, minimal overhead on the links at the data plane. Note that most of the fuzz traffic is dropped at the first switch unless there is a flow rule to match that traffic and thus, incurring even less overhead on the links.
\vspace{-.75em}
\subsection{\dpl Overhead} 
\label{sec:I6}
We generated different packet sizes from 64 bytes to 1500 bytes at almost Gbps rate on the switches running the \dpl software of \system and the native OvS switches. We added flow rules on our switches to match the packets and tag them by the \prot shim header using our \textit{push\_verify}, \textit{set\_\vr} and \textit{set\_\vp} actions in the flow rule. Then, we measured the average throughput over 10 runs. 

We observe that the \prot shim header and the tagging mechanism incurred negligible throughput drop as compared to the native OvS. However, a throughput drop of 1.1.\% is observed only on the entry switch/exit switch as \textit{push\_verify} actions happen on the entry switch/exit switch to insert the \prot shim header. Furthermore, sFlow sampling is done at the exit switch only. On the contrary, \svp and \svr just set the \vp and \vr fields respectively and have no observable impact. Thus, \system introduces minimal packet processing overheads atop OvS. 
\vspace{-.5em}
\section{Related Work}
\label{sec:rel}
In addition to the related work covered in~\cref{sec:intro} that includes the existing literature based on~\cite{Heller2013} and Table~\ref{tab:rel}, we now will navigate the landscape of related works and compare them to \system in terms of the Type-p and Type-a faults which cause inconsistency (\cref{sec:un}). The related work in the area of control plane~\cite{fayaz2016buzz,Mai2011, Kazemian2012, Kazemian2013, veriflow, Lopes2015, Yang2016,Div, Canini2012,troubleshooting,batfish} either check the controller-applications or the control-plane compliance with the high-level network policy. These approaches are insufficient to check the physical data plane compliance with the control plane. As illustrated in Table~\ref{tab:rel}, we navigate the landscape of the data plane approaches and compare them with \system based on the ability to detect Type-p and Type-a faults. It is worth noting that the approaches either test the rules or the paths whereas \system tests both together. In the case of Type-p match faults (\cref{sec:act}) when the path is same even if different rule is matched, path trajectory tools~\cite{Zhang2016,Handigol2014,Narayana2016,Tamanna2015,pathdump,Agarwal2014,foces} fail. The approaches based on active-probing~\cite{Zeng2014,perevsini2015monocle,Agarwal2014,zhang2017,Rulescope,sdnprobe} do not detect the Type-a faults (\cref{sec:2}) caused by hidden or misconfigured rules on the data plane which only match the fuzz traffic. These tools, however, only generate the probes to test the rules known or synced to the controller. Such Type-a faults are detected by \system. Recently, Shukla et al. developed P4CONSIST~\cite{p4consist} which relies on production traffic to detect path-based control-data plane inconsistency in P4 SDNs. However, it cannot localize and calso, cannot detect rule-based control-data plane inconsistency in SDNs like~\system. Furthermore, the packet header space coverage is less than~\system as it solely relies on production traffic. Recently, P4RL~\cite{shukla2} leveraged machine learning-guided fuzzing to detect path deviations in P4 switches. However, P4RL is limited to P4 switches only.  
\vspace{-.65em}
\section{Discussion: How far to go?}
\label{sec:discussion}
\begin{figure}[t!]
	\vspace{-.75em}
	\centering
	\includegraphics[width=0.95\columnwidth]{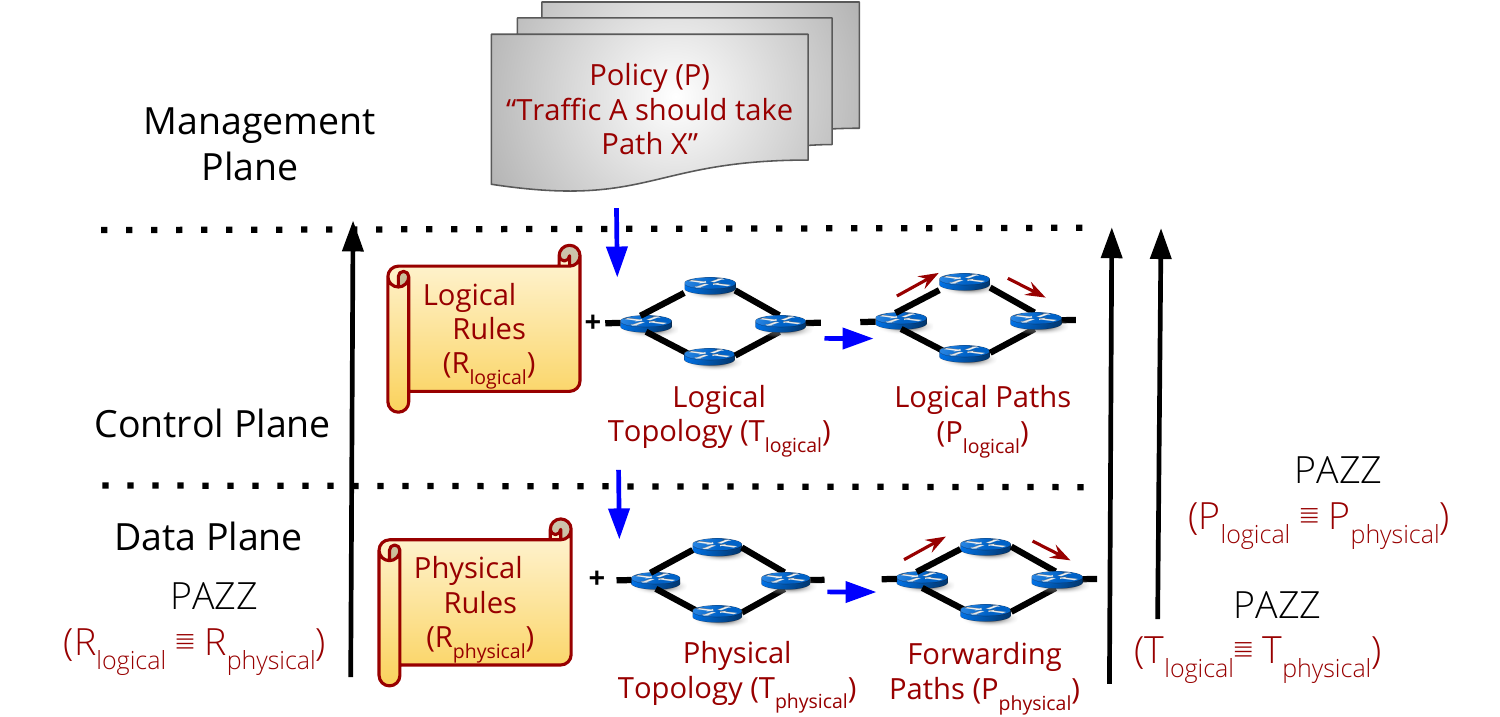}
	\captionof{figure}{\system provides coverage on the forwarding flow rules, topology and paths.\label{fig:Pazz}}
	\vspace{-.5em}
\end{figure}
\vspace{-.25em}
Motivated by our analysis of existing faults which cause inconsistency in SDNs
and the fact that network consistency is only as good as its weakest
link, we have argued that we need to take consistency checks further.
Using an analogy to program testing, we proposed a novel ``bottom-up''
methodology, making the case for network state fuzzing.

However, the question of ``how far to go?'' in terms of consistency checks
is a non-trivial one and needs further attention.
In particular, while methodologies such as \system
provide a framework to systematically find (and subsequently remove)
faults, \emph{in theory}, they raise a question of \emph{how much
	resources to invest}. Indeed, we observe that providing full coverage of forwarding flow rules, topology, and forwarding paths is inherently expensive:
Consider a simplistic linear network of $n=4$ switches with two links between each switch.
Assume IP traffic enters from a source and exits on a destination, and each switch
stores a simple forwarding rule: switch $S_i$ for $i = \{1,2,3...n\}$ 
bases its forwarding decisions solely on the 
$i^\textit{th}$ least significant bit of the destination IP address:
packets of even bits are forwarded to the upper link (\emph{up}) and packets of
odd bits to the lower (\emph{lo}) link. While such a configuration requires only
a small constant number of rules per switch, the number of possible paths is \emph{exponential}
in $n$: the set of paths is given by $\{\emph{up},\emph{lo}\}^{n-1}$. 
In addition to the combinatorial complexity of testing static configurations, 
we face the challenge that both the (distributed) control plane and the data plane state
(as well as possibly the network connecting the controllers to the data plane elements) 
evolve over time, possibly in an asynchronous (or even stateful~\cite{fayaz2016buzz}) 
and hard to predict manner. 
This temporal dimension exacerbates the challenge of identifying faults,
raising the question ``how \emph{frequently} to check''.

Finally, the question of finding faults can go beyond
deciding on the right amount of resources to invest:
as any traffic-based testing introduces 
the inherent ``Schr\"odinger's cat'' 
problem that tagged traffic may not be subject to  the 
same treatment as the untagged one.

We note that current trends towards programmable data planes~\cite{Bosshart2014} and hardware~\cite{firestone2018azure} may
aid in solving some of the issues this paper raises. INT~\cite{kim2015band} in P4 specifically tries to solve the problem of network monitoring but is yet to be made scalable for the whole network and is limited to the P4 switches.  Still, leveraging programmability,
future SDNs will encompass even more possibilities of faults with a
mix of vendor-code, reusable libraries, and in-house code. Yet, there will no
longer be a common API, i.e., OpenFlow.  As such the general problem will
persist and we will have to explore how to extend the \system methodology to programmable networks.

Overall, as illustrated in Figure~\ref{fig:Pazz}, \system checks consistency at all levels between control and the data plane, i.e., $P$\textsubscript{logical} $\equiv$ $P$\textsubscript{physical}, $T$\textsubscript{logical} $\equiv$ $T$\textsubscript{physical}, and $R$\textsubscript{logical} $\equiv$ $R$\textsubscript{physical}.
\vspace{-1.2em}
\section{Conclusion \& Future Work}
\label{sec:Conclusion}
This paper presented \system, a novel network verification methodology that automatically detects and localizes the data plane faults manifesting as inconsistency in SDNs. \system periodically fuzz tests the packet header space and compares the expected control plane state with the actual data plane state. The tagging mechanism tracks the paths and rules at the data plane while the reachability graph at the control plane tracks paths and rules to help \system in verifying consistency. Our evaluation of \system over real network topologies and configurations showed that \system efficiently detects and localizes the faults causing inconsistency while requiring minimal data plane resources (e.g., switch rules and link bandwidth).

Currently, we do not handle transient faults~\cite{shukla2016towards}, packet rewriting and checking of backup rules lays the groundwork for our future work. For improving the performance of \system further, we consider multithreading at the \consistencychecker will allow multiple flows to be checked in parallel. For \dpl, Data Plane Development Kit (DPDK) or P4 can improve the \dpl performance. We will consider performance-related faults like throughput, latency and packet-loss. Distributed SDN controller and P4-specific hardware~\cite{Bosshart2014,firestone2018azure} design and implementation will also be a part of our future work.
\vspace{-1em}
\bibliographystyle{IEEEtran}
\bibliography{IEEEabrv,literature}
\end{document}